\documentclass[journal,twoside]{IEEEtran}
\usepackage[T1]{fontenc}
\usepackage{cite}
\usepackage{amsmath,amssymb,amsfonts}
\usepackage{algorithmic}
\usepackage[]{graphicx}
\usepackage[mathlines,switch]{lineno}
\usepackage{multicol,multirow}
\usepackage{epsfig}
\usepackage{TUSON}
\makeatletter
\renewenvironment{abstract}{\global\setbox0=\vbox\bgroup\oabstract}{\oendabstract\egroup\box0}
\makeatother
\usepackage{array}
\usepackage{threeparttable}
\usepackage{upgreek}
\usepackage{changes}

\title{Spurious-Free Lithium Niobate Bulk Acoustic Wave Resonator with Grounded-Ring Electrode}
\author{{Vakhtang Chulukhadze, Kristi Nguyen, Eric Stolt, Kilian Shambaugh, Weston Braun, Tzu-Hsuan Hsu, Osama Jameel, Juan Rivas-Davila, and Ruochen Lu}
\thanks{This paper is an expanded version of the IEEE International Frequency Control Symposium (IFCS) 2023.}
\thanks{This work was supported by Defense Advanced Research Projects Agency (DARPA) High Operational Temperature Sensors (HOTS) program, and National Science Foundation (NSF) ECCS Award No. 2530883.}
\thanks{Vakhtang Chulukhadze, Kristi Nguyen, Tzu-Hsuan Hsu, and Ruochen Lu are with the Department of Electrical and Computer Engineering at the University of Texas at Austin. (e-mail: vatoc@utexas.edu). Eric Stolt, Weston Braun, and Juan Rivas-Davila are with the Department of Electrical Engineering at Stanford University. Kilian Shambaugh and Osama Jameel are with Polytec GMBH.}}

\IEEEaftertitletext{\GA{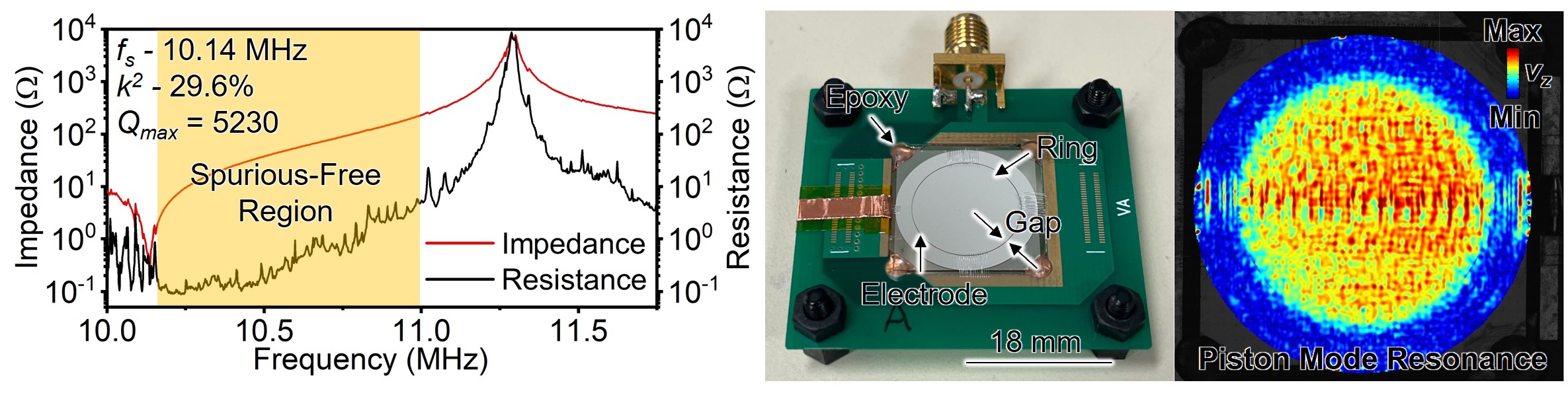}
\Abstract{\begin{abstract}
High-performance piezoelectric resonators are promising energy-storage elements for piezoelectric power conversion because of their compact footprint and low loss at frequencies where conventional magnetic components become bulky and inefficient. However, their practical use is often limited by the trade-off between a high electromechanical coupling coefficient (\(k^2\)) for wide-band operation and the emergence of spurious acoustic modes that limit the resonators' inductive bandwidth. This work reports a spurious-free thickness-extensional (TE)-mode bulk acoustic wave (BAW) resonator in single-crystal lithium niobate (LN) based on a grounded-ring electrode architecture. The proposed structure is analyzed through simulation and experimentally validated using electrical characterization and laser Doppler vibrometry (LDV). The results show that the grounded ring modifies the effective boundary conditions of the acoustic device, enabling a piston-like modal response that suppresses lateral spurious modes across the inductive band. The demonstrated device operates at 10.14 MHz and achieves an electromechanical coupling coefficient of 29.6\%, a maximum in-band Bode quality factor (\(Q_\textit{Bode}\)) of 5230, and a figure of merit (FoM, $Q\cdot k^2$) of 1548. These results establish the grounded-ring TE-mode LN BAW resonator as a practical platform for piezoelectric power conversion and a broader design approach for realizing high-performance spurious-free acoustic resonators.
\hblne
\end{abstract}}
\vspace{1\baselineskip}\vspace*{-1pt}
}

\begin{document}

\bstctlcite{IEEEexample:BSTcontrol}

\maketitle

\Highlights[1]{A grounded-ring electrode enables a spurious-free thickness-extensional (TE)-mode lithium niobate (LN) bulk acoustic wave (BAW) resonator, validated by simulation, electrical characterization, and laser Doppler vibrometry (LDV).}

\Highlights[2]{The proposed structure suppresses lateral spurious modes by altering the effective boundary conditions and exciting a piston-like modal response across the inductive band.}

\Highlights[3]{The device demonstrates spurious-free operation at 10.14 MHz with an in-band Bode quality factor (\(Q_\textit{Bode}\)) of 5230, an electromechanical coupling coefficient \(k^2\) of 29.6\%, and a figure of merit (FoM) of 1548, providing a practical path toward high-performance acoustic resonators for power conversion and related applications.}


\PrintHighlights

\section{Introduction}
\label{sec:introduction}
\IEEEPARstart{M}{odern} power electronic systems have continuously shown the need for compact power conversion with a high power density. Operating direct current-to-direct current (DC-DC) converters at higher switching frequencies is a promising approach in this regard \cite{perreault_opportunities_2009}. However, conventional power-conversion technologies have been unable to accommodate such demand due to the size and performance constraints imposed by magnetic inductors at higher frequencies \cite{sullivan_size_2016, hanson_measurements_2016}. Conversely, piezoelectric acoustic resonators, which have long been established as radio frequency (RF) filtering solutions with compactness and low-loss, offer intrinsic advantages over their electromagnetic (EM) counterparts while showing significant inductive behavior \cite{gong_microwave_2021, lu_rf_2021, lu_recent_2025}. These advantages have positioned piezoelectric devices as compelling alternatives to bulky inductors in DC-DC converters \cite{boles_piezoelectric-based_2022}.

Piezoelectric resonators have been readily integrated into DC-DC power-conversion circuits, leveraging a cascade of zero-voltage-switching sequences to drive an alternating-current (AC) oscillation in the acoustic resonator \cite{braun_inductorless_2020, braun_optimized_2021, touhami_piezoelectric_2021, boles_dcdc_2022, boles_piezoelectric-resonator-based_2023}. These systems have demonstrated promising results, yielding compact power converters with high power densities and efficiency. However, they typically exhibit a limited voltage conversion ratio because unwanted resonances are excited within the resonator's inductive bandwidth. These spurious acoustic modes are a byproduct of piezoelectric platforms with high electromechanical efficiency (\(k^2\)), and have been the primary limiting factor in piezoelectric power converter performance, especially at high frequencies \cite{stolt_fixed-frequency_2021}. This work presents an acoustic design framework that overcomes this intrinsic limitation, yielding a high-performance, spurious-free acoustic resonator for piezoelectric power conversion.

\begin{figure}[!t]
\centerline{
\includegraphics[width=\linewidth]{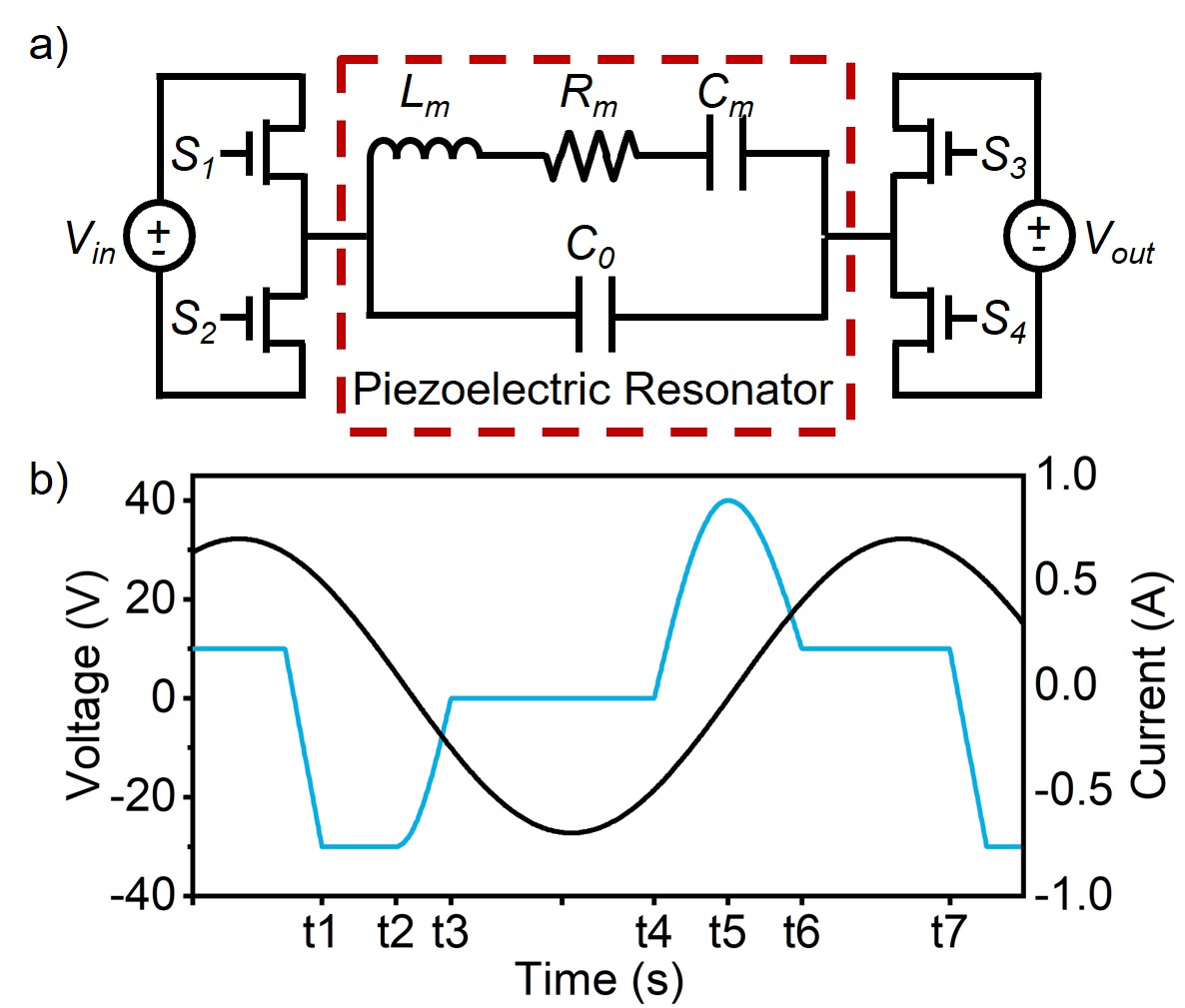}}
\caption{(a) BVD circuit model of piezoelectric resonator integrated into a power converter. (b) Idealized voltage and current waveforms for 40~V input, 30~V output \cite{stolt_fixed-frequency_2021}.}
\label{circuit_model}
\end{figure}

A piezoelectric power converting circuit can be modeled as a resonator connected to various switch configurations (\(S_1, S_2, S_3, S_4\)) and direct current (DC) voltage sources (\( V_\text{in}, V_\text{out}\)) [Fig. \ref{circuit_model} (a)]. The resonator itself is modeled with the Butterworth-Van-Dyke (BVD) equivalent circuit model, consisting of a series motional inductor (\(L_m\)), resistor (\(R_m\)), and capacitor (\( C_m \)) connected in parallel with a static capacitance (\(C_0\)), making the acoustic resonator the sole energy-storage element in the power-conversion circuit \cite{van_dyke_piezo-electric_1928}. To mimic the operation of a magnetic inductor, the resonator must be operated between its resonance (\(f_s\)) and anti-resonance (\(f_p\)) frequencies. Within this band, the acoustic device shows net inductive behavior. Maximum efficiency is achieved near \(f_s\) and is highly influenced by the associated quality factor (\(Q\)) \cite{daniel_nonlinear_2024}. The efficiency then decreases as the operating frequency approaches \(f_p\). The frequency band between the two is determined by the resonator's \(k^2\). As a result, the product between these two metrics is utilized as the figure of merit (FoM, $Q\cdot k^2$). In addition to \(Q\) and \(k^2\), the emergence of unwanted in-band spurious modes further limits the performance of acoustic resonators for power conversion. Spurious modes introduce sharp spikes in the resonator's impedance response, causing high power-converter losses and limiting the effective operational bandwidth. Consequently, an acoustic resonator design that exhibits high \(Q\) and \(k^2\) and a spurious-free impedance response will be indispensable for developing piezoelectric power conversion. To achieve this goal, careful material and acoustic mode selection are key design criteria.

\newcolumntype{M}[1]{>{\centering\arraybackslash}m{#1}}

\begin{table}[!t]
\caption{SOA of Piezoelectric Power Converter Bulk Resonators}
\label{tab1}
\begin{threeparttable}
\centering
\renewcommand{\arraystretch}{1.2}
\setlength{\tabcolsep}{3.5pt}
\begin{tabular}{M{70pt}
                M{25pt}
                M{25pt}
                M{25pt}
                M{25pt}
                M{40pt}}
\hline\hline
Reference & $f_s$ (MHz) & $k^2$ & $Q$* & FoM & Spurious Suppression \\
\hline
PZT-Radial\cite{boles_dcdc_2022} & 0.48 & 19\% & 1030 & 196 & Y \\
PZT-Radial\cite{daniel_nonlinear_2024} & 0.17 & 42.5\% & 1198 & 509 & N  \\
PZT-TE\cite{boles_evaluating_2022} & 0.61 & 31\% & 2500 & 775 & N  \\
LN-TS\cite{wu_spurious-free_2022} & 3.55 & 53\% & N/A & N/A & N \\
LN-TS\cite{nguyen_near-spurious-free_2022} & 5.94 & 45\% & 3500 & 1575 & Y  \\
LN-TE\cite{touhami_piezoelectric_2021}  & 6.28 & 25.5\% & 3700 & 944 & N \\
LN-TE\cite{braun_optimized_2021}  & 6.82 & 29\% & 4178 & 1212 & Y  \\
LN-Radial\cite{stolt_stacked_2024}  & 0.37 & 23.4\% & 17000 & 4000 & N  \\
P3F LN-TE\cite{yao_periodically_2025} & 19.23 & 29\% & 3187 & 928 & N  \\
LT-TE\cite{yao_lithium_2025}  & 6.50 & 8.8\% & 1698 & 149 & Y  \\
LN-TE\cite{chulukhadze_lithium_2024}  & 10.13 & 29.1\% & 2128 & 620 & Y  \\
AlN-TE\cite{yao_single-crystal_2026}  & 13.52 & 6.1\% & 1677 & 103 & Y  \\
\textbf{LN-TE (This Work)} & \textbf{10.14} & \textbf{29.6\%} & \textbf{5230} & \textbf{1548} & \textbf{Y}  \\
\hline\hline
\end{tabular}
\begin{tablenotes}
\footnotesize
\item[] \centering *For each reference, the maximum reported \(Q\) value was used.
\end{tablenotes}
\end{threeparttable}
\end{table}

Piezoelectric power converters have predominantly utilized lead zirconate titanate (PZT) and lithium niobate (LN) platforms due to their high figures-of-merit (FoMs) across various acoustic modes \cite{touhami_piezoelectric_2021, braun_optimized_2021}. PZT-based acoustic resonators have demonstrated high power densities in low-frequency applications using off-the-shelf components. However, their performance is often constrained by the high relative permittivity (\(\varepsilon_r\)) and the inherent nonlinearity at high applied electrical fields \cite{daniel_nonlinear_2024}. In contrast, single-crystal LN possesses \(k^2\) comparable to PZT, but exhibits a significantly lower \(\varepsilon_r\) and superior linearity. These characteristics enable compact and efficient operation in high-frequency and high-power regimes. This contrast is reflected in Table \ref{tab1} \cite{boles_dcdc_2022, daniel_nonlinear_2024, boles_evaluating_2022,wu_spurious-free_2022, nguyen_near-spurious-free_2022, touhami_piezoelectric_2021, braun_optimized_2021, stolt_stacked_2024, yao_periodically_2025, yao_lithium_2025, chulukhadze_lithium_2024, yao_single-crystal_2026}, which shows that the state-of-the-art (SoA) for sub-MHz power conversion consists primarily of radial-mode PZT devices. Conversely, LN-based resonators dominate at higher frequencies, typically operating in bulk acoustic wave (BAW) modes, such as thickness-shear (TS) and thickness-extensional (TE) \cite{nguyen_near-spurious-free_2022, braun_optimized_2021, yao_periodically_2025}. Related materials, such as lithium tantalate (LT) and aluminum nitride (AlN), also show significant promise in their temperature stability, despite the lower \(k^2\) \cite{yao_lithium_2025, yao_single-crystal_2026}. However, regardless of their superior FoMs, existing LN platforms suffer from a severely limited operational bandwidth due to in-band spurious acoustic modes \cite{wu_spurious-free_2022, yang_lateral_2021}. These unwanted resonances, especially near \(f_s\), introduce regions of resistive loss for the piezoelectric power converter, thereby degrading converter efficiency and limiting the voltage conversion ratio \cite{braun_optimized_2021}. While solutions, e.g., active-region electrodes, have been proposed, they cannot overcome this shortcoming without compromising the material's performance \cite{chulukhadze_lithium_2024}.

This work addresses these fundamental limitations by implementing a novel spurious-free acoustic device design for high-frequency piezoelectric power conversion. Wideband performance suitable for a power converter was achieved using a BAW resonator operating in TE mode, with a grounded-ring structure to facilitate spurious-mode suppression. As a result, we demonstrate a spurious-free BAW resonator operating in the TE mode at 10.14 MHz with a high \(Q\) of 5230, and a high \(k^2\) of 29.6\%, resulting in an FoM of 1548. This paper substantially extends the conference report in \cite{nguyen_spurious-free_2023} with three key additions: (1) full LDV characterization of all three device variants, (2) a physically motivated analysis of the spurious-mode suppression mechanism based on Lamb wave dispersion, and (3) geometric design rules linking gap width and ring width to the plate's dispersion characteristics. The proposed resonator has been integrated into a 3.2~kW DC-DC converter in \cite{stolt_spurious-free_2024}, but the acoustic resonator design and its operating mechanism have not been previously discussed.

\section{Design and Simulation}
\label{sec:design}

\begin{figure}[!t]
\centerline{\includegraphics[width=\columnwidth]{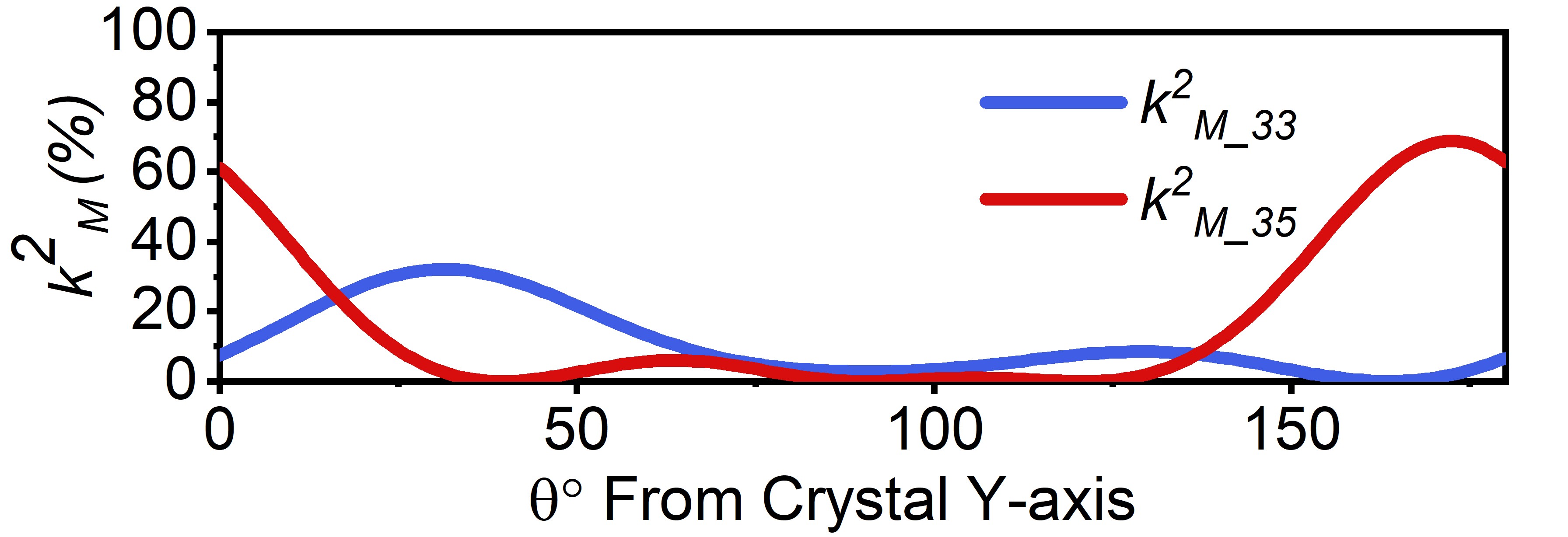}}
\caption{Coupling coefficients as a function of rotation from the crystal Y-axis, for TS mode ($ k_\textit{M\_35}^2$) and TE mode ($ k_\textit{M\_33}^2$). 36Y-cut LN combines high $k_\textit{M\_33}^2$ with low parasitic $k_\textit{M\_35}^2$.}
\label{GroundedRing_orientation}
\end{figure}

\subsection{Mode and Orientation Selection}

LN exhibits a densely populated electromechanical coupling tensor, enabling efficient piezoelectric excitation of a wide range of shear and compressional acoustic waves. Previous studies have examined many of their modal profiles and key design parameters to assess their suitability for power conversion. While radial-mode LN devices are promising for kHz operation, TS and TE modes are the primary choice for high-frequency power conversion at MHz and beyond \cite{boles_evaluating_2022}.

\begin{figure}[!t]
\centerline{\includegraphics[width=\columnwidth]{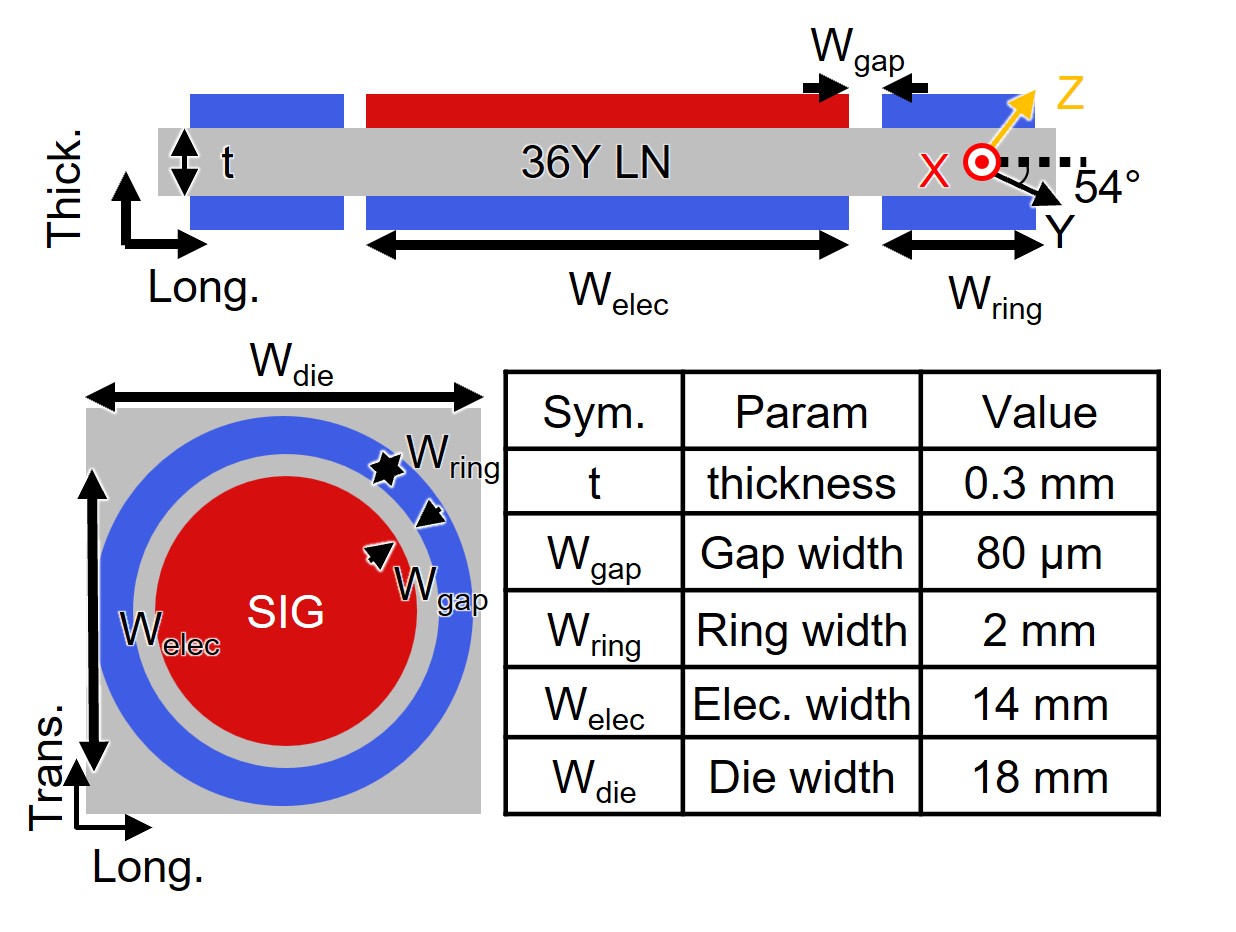}}
\caption{Illustration of cross-section and top view of the proposed resonator design with a grounded ring, with parameters tabulated. All electrodes are 300 nm thick, significantly thinner than LN.}
\label{GroundedRing_schematic}
\end{figure}

Recent demonstrations of high FoM TS-mode resonators in the MHz range have highlighted challenges in maintaining a spurious-free response within the inductive bandwidth \cite{nguyen_spurious-free_2023}. While existing suppression methods can tune the frequency of lateral TS-mode overtones by varying the electrode duty cycle or wavelength, they do not provide broadband suppression of spurious modes. In contrast, TE-mode devices benefit from dispersion characteristics that enable spurious-free acoustic resonators in RF applications \cite{thalhammer_4e-3_2006}. Consequently, the TE-mode was adopted to realize a spurious-free, high FoM acoustic resonator for piezoelectric power conversion.

To maximize TE-mode transduction efficiency while minimizing parasitic coupling, the strong anisotropy of LN must be carefully characterized. TE mode devices leverage a thickness-oriented electrical field to drive vibrations using the \(e_\textit{33}\) electromechanical coupling tensor coefficient. Consequently, the other thickness-field tensor components must be minimized to avoid piezoelectric coupling with unwanted shear modes. The selection process was guided by the material \(k^2\) ($k_\textit{M\_ij}^2$), which is evaluated assuming the entire piezoelectric volume participates in the vibration, and shows the material's maximum piezoelectric transduction efficiency. The specific modal profile is selected using the subscripts, where $i$ is the electrical field direction, and $j$ is the stress component, as follows:
\begin{equation}
 k_\textit{M\_ij}^2  = \frac{e_\textit{ij}^2}{\varepsilon^S_\textit{ii}\cdot c^E_\textit{jj}}
\end{equation}
where the subscript $M$ denotes the material coupling coefficient (assuming full piezoelectric volume participation). Among commercially available LN wafers, rotated Y-cut crystals tend to exhibit higher TE efficiency. To identify the optimal cut, we plot $k_\textit{M\_ij}^2$ for the TS and TE modes as a function of the rotational angle \(\theta\) in \(ZXZ\) Euler angle rotation (0, 90 - \(\theta\), 0), as shown in Fig. \ref{GroundedRing_orientation}.  Accordingly, 36Y LN was chosen for its high $ k_\textit{M\_33}^2$ paired with a low parasitic $k_\textit{M\_35}^2$. The remaining key step was to tailor the acoustic design to suppress TE spurious modes in 36Y LN.

\subsection{Grounded Ring for Spurious-Mode Suppression}

\begin{figure}[!t]
\centerline{\includegraphics[width=\columnwidth]{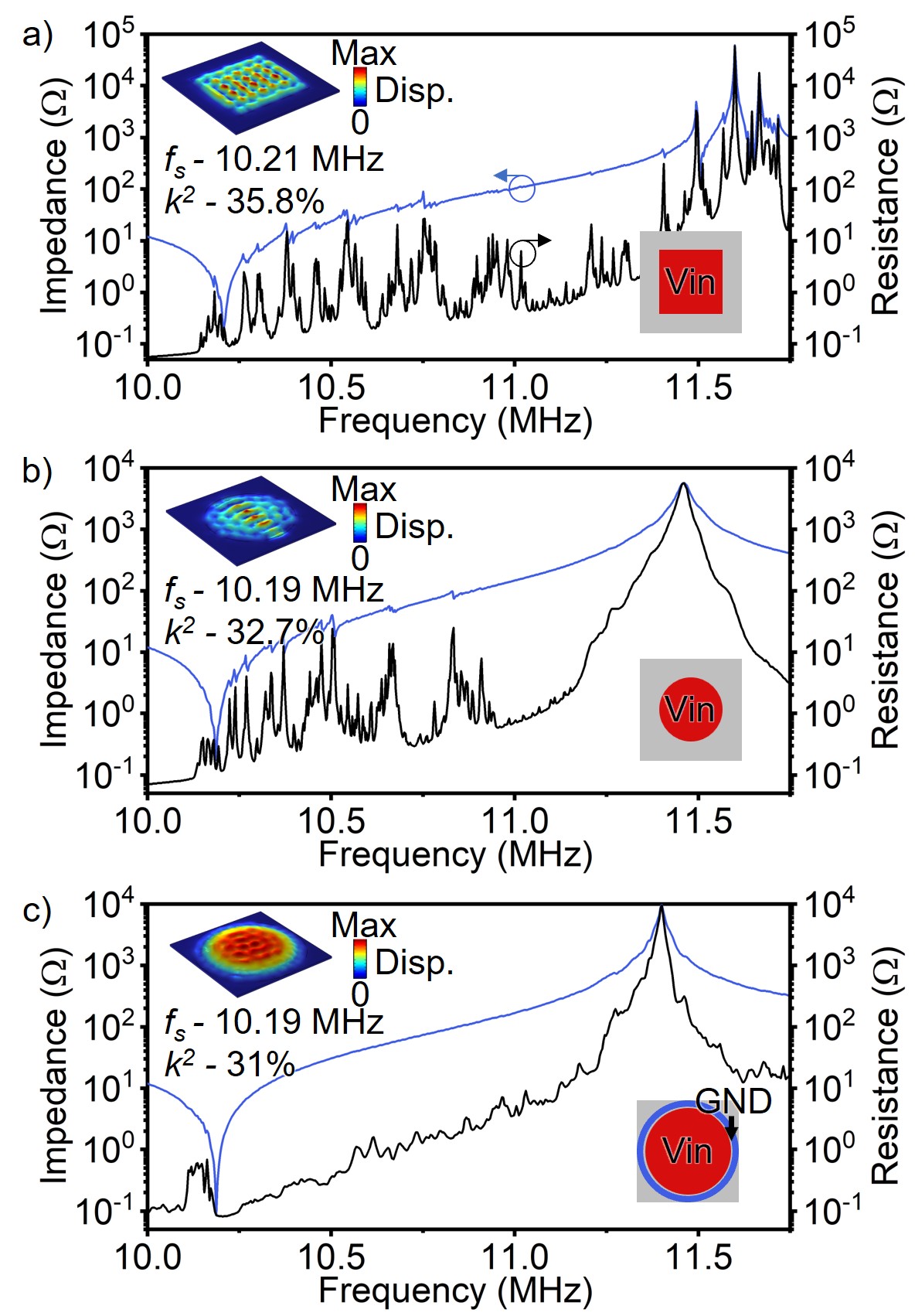}}
\caption{Simulated impedance and resistance for (a) rectangular, (b) circular, and (c) grounded-ring TE resonators. Inset images show 3D mode shapes and electrode configurations.}
\label{GroundedRing_ImpedanceSims}
\end{figure}

The mechanism for spurious-mode suppression is based on a novel electrode design, as illustrated in Fig. \ref{GroundedRing_schematic}. The structure comprises three distinct regions: an active area defined by a signal-top and a ground-bottom electrode, a non-metalized sub-wavelength gap, and a metalized ring region surrounding the active area, connected to the ground on both the top and bottom. The specific dimensions are provided in the inset table of Fig. \ref{GroundedRing_schematic}, highlighting the gap's small size relative to the other resonator dimensions.

To evaluate the proposed structure, a 300 \(\upmu\)m-thick 36Y LN bulk wafer was modeled to target a TE-mode resonance at 10 MHz [Fig. \ref{GroundedRing_ImpedanceSims}(c)]. The reference designs were sized to match the active area of the grounded-ring structure. The rectangular reference is a fully metalized 14$\times$14~mm$^2$ square [Fig.~\ref{GroundedRing_ImpedanceSims}(a)], and the circular reference is a fully metalized disk with a 7~mm radius [Fig.~\ref{GroundedRing_ImpedanceSims}(b)].

The proposed circular grounded-ring resonator design was benchmarked against conventional rectangular and circular geometries using 3D COMSOL finite element analysis (FEA), with the resulting impedance and resistance spectra seen in Fig. \ref{GroundedRing_ImpedanceSims} (a)-(c). The simulations assume a $Q$ of 2000, consistent with experimentally observed values for single-crystal LN resonators in the low-MHz regime \cite{colombo_high-figure--merit_2020}.

\begin{figure}[!t]
\centerline{\includegraphics[width=\columnwidth]{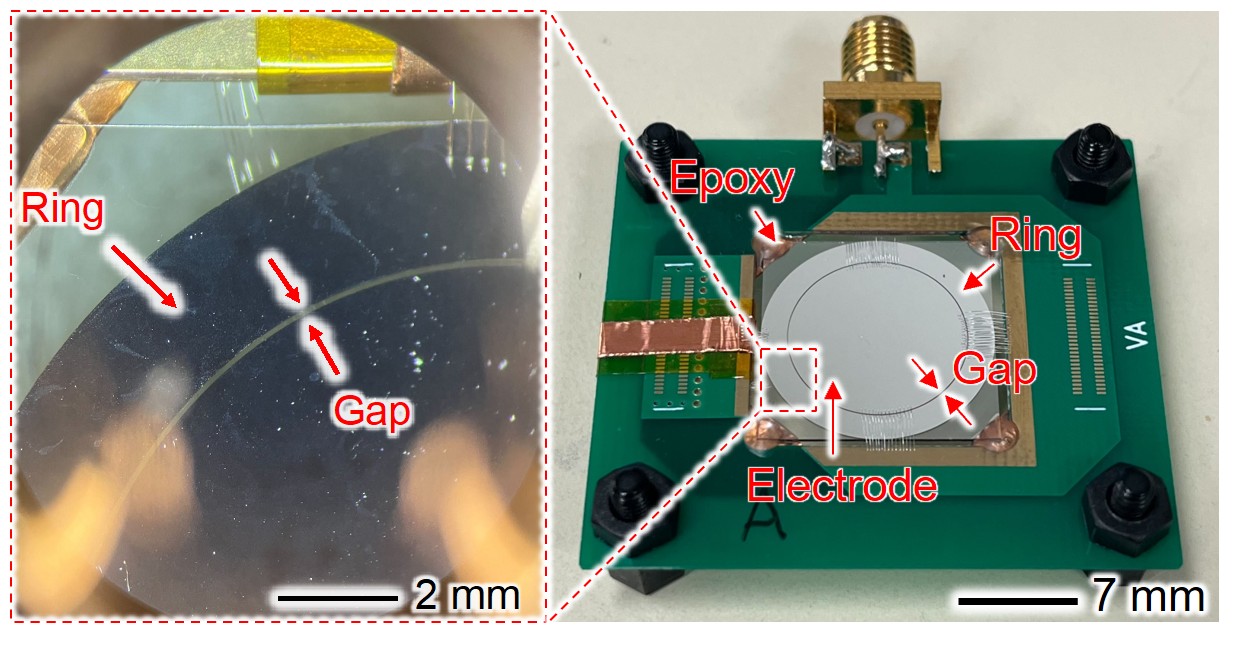}}
\caption{Optical image of fabricated LN resonator mounted on a PCB. The inset image shows a magnified view of the ring and gap regions.}
\label{mounteddevice}
\end{figure}

Conventional rectangular and circular TE-mode designs [Fig. \ref{GroundedRing_ImpedanceSims} (a) and (b)] exhibit a high parasitic modal density within the resonator's inductive bandwidth. Spurious modes are especially dense near \(f_s\), where the power converters operate. These unwanted resonances are lateral overtones of the TE mode, manifesting as sharp spikes in resistance that contribute to significant converter losses. The reference designs yield an approximate resonance frequency of 10.2 MHz, and their \(k^2\) is evaluated using the standard BVD relation \cite{van_dyke_piezo-electric_1928}:
\begin{equation}
 k^2  = \frac{\pi^2}{8}((\frac{f_p}{f_s})^2-1)
 \label{eqk2}
\end{equation}
The rectangular and circular reference designs yield \(k^2\) of 35.8\% and 32.7\%, respectively. Although, spurious modes near $f_p$ obscure the true anti-resonance, causing the BVD-extracted $f_p/f_s$ ratio in (\ref{eqk2}) to overestimate $k^2$ \cite{lu_accurate_2019}.

In contrast, the proposed circular grounded-ring TE resonator [Fig. \ref{GroundedRing_ImpedanceSims} (c)] achieves a smooth resistance curve, effectively eliminating in-band spurious modes. Unlike the response seen in Fig. \ref{GroundedRing_ImpedanceSims} (a) and (b), the spurious-free nature of the proposed structure allows for a straightforward extraction of \(k^2\) of 31\%. The achieved spurious-free performance is consistent with the calculated \(k^2_\textit{M}\), and close to those extracted from the conventional designs, indicating that the suppression mechanism does not compromise the electromechanical efficiency.

Compared with prior spurious-free TE-mode resonators, which require regions with three distinct dispersion characteristics via additional metal or dielectric deposition or etching, this design significantly simplifies fabrication by requiring only two lithographically defined electrode regions (active electrode and a grounded ring) on each wafer face, patterned in a single metal deposition and liftoff step \cite{thalhammer_4e-3_2006}.

\section{Experimental Results}

\subsection{Device Fabrication}

\begin{figure}[!t]
\centerline{\includegraphics[width=\columnwidth]{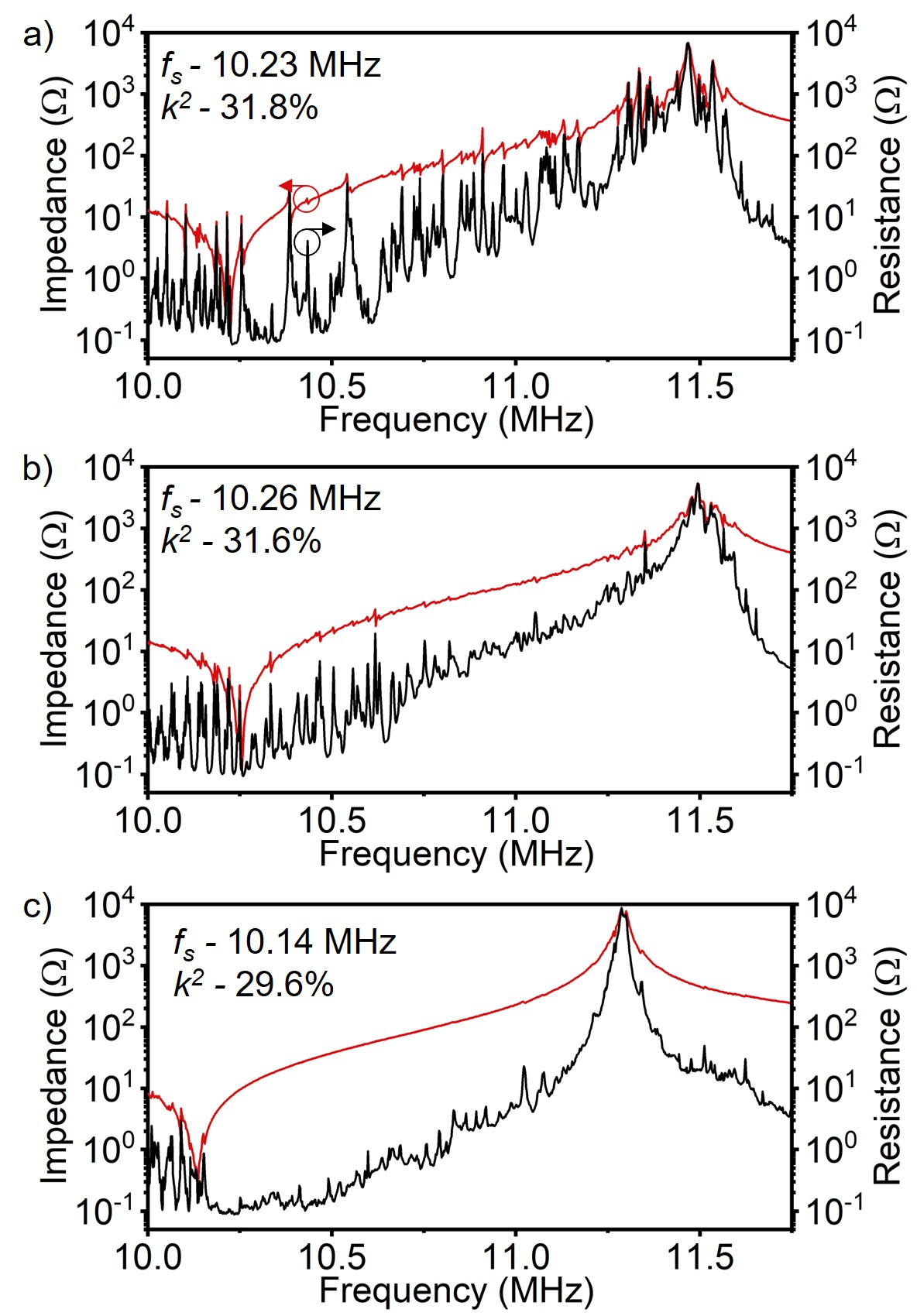}}
\caption{Impedance and resistance of resonators with (a) a conventional rectangular design, (b) a conventional circular design, and (c) the proposed grounded-ring design. The grounded-ring resonator shows spurious-mode suppression and a high \(k^2\).}
\label{impedancemeasurements}
\end{figure}

The design seen in Fig. \ref{GroundedRing_schematic} was implemented using a bulk 0.3~mm thick, 4-inch congruent 36Y-cut LN wafer. The fabrication process began with metal deposition on the front and back sides of the wafer using standard photolithography tools and electron-beam evaporation. 300 nm of Au was chosen as the electrode material and thickness due to its excellent electrical conductivity and its ability to enable stable wire-bond interconnects. At this thickness, the Au electrode mass loading on the 300~$\upmu$m LN plate has a negligible effect on the resonance frequency. Following metalization, the wafer was diced into 18$\times$18 mm\(^2\) individual dies and mounted onto a printed circuit board (PCB) for measurement, as shown in Fig. \ref{mounteddevice}. The mounting was performed by epoxying the corners of the fabricated die to the PCB. At the same time, an electrical connection was established via wire bonding, with the signal and ground routed to an SMA connector for characterization. With a starting resonator die size of 18$\times$18 mm\(^2\), the mounted device has an area of 28$\times$28 mm\(^2\), highlighting the compactness of the proposed solution relative to its EM counterparts.

\subsection{Electrical Measurements}

The fabricated device was measured using a vector network analyzer (VNA) in accordance with standard calibration procedures. The results are shown in Fig. \ref{impedancemeasurements} (a)-(c), closely matching the simulated results in Fig. \ref{GroundedRing_ImpedanceSims}. As in the simulation, the conventional designs [Fig. \ref{impedancemeasurements} (a) and (b)] exhibit significant spurious modes, leading to an exaggerated \(k^2\) of 31.8\% and 31.6\% for the rectangular and the circular reference devices, respectively \cite{lu_accurate_2019}. In contrast, the grounded-ring design exhibits a spurious-free response while yielding a \(k^2\) of 29.6\%, which closely matches the simulation and indicates minimal loss in performance. Minor deviations in the measured \(f_s\) across the fabricated devices are attributed to inherent in-plane thickness non-uniformity in the LN wafer.

\begin{figure}[!t]
\centerline{\includegraphics[width=\columnwidth]{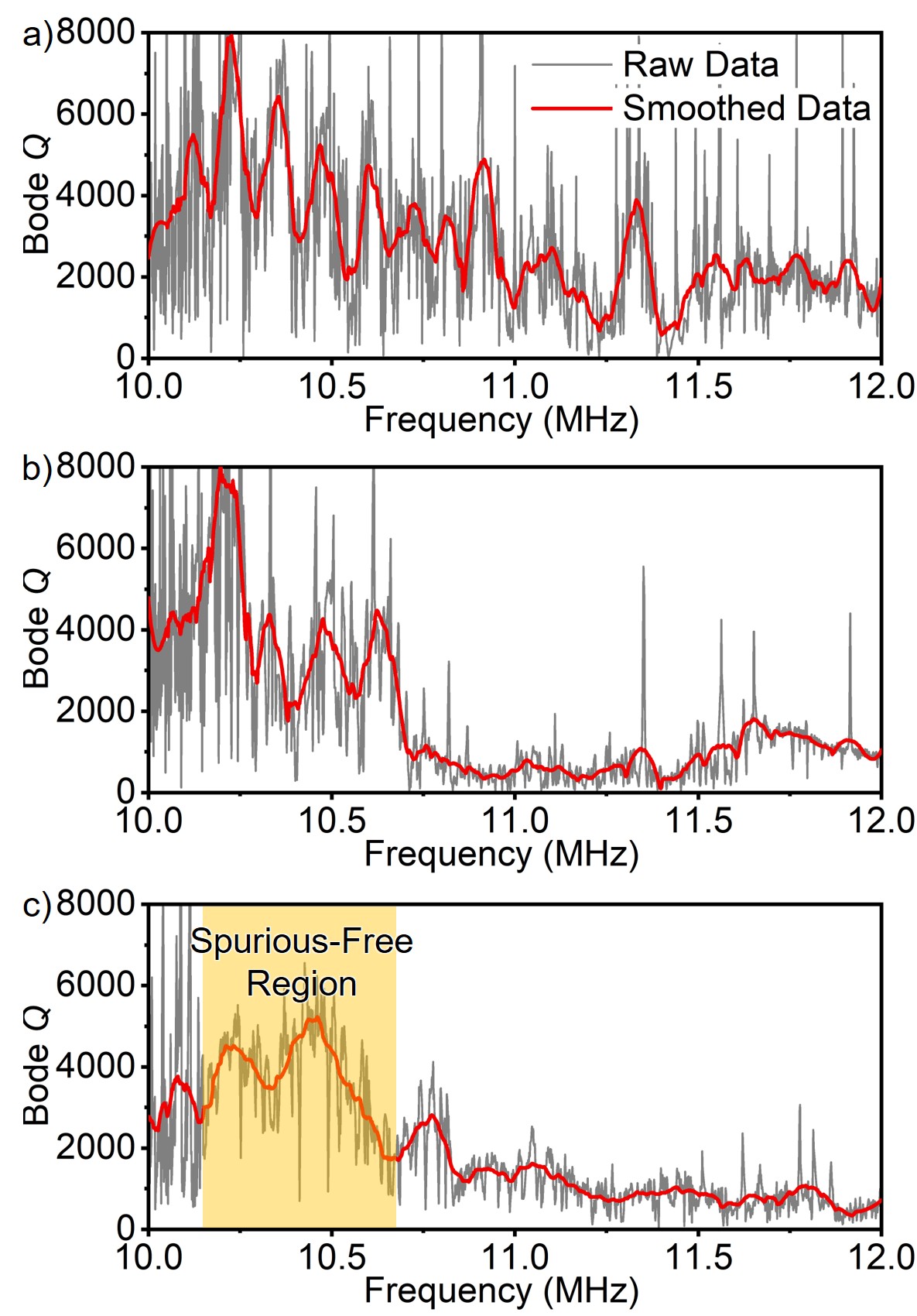}}
\caption{Measured Bode \(Q\) from (a) a conventional rectangular device, (b) a conventional circular device, (c) the proposed grounded-ring design. The results highlight high \(Q_\textit{Bode}\) achieved within the spurious-free band of the grounded-ring device. }
\label{BodeQ}
\end{figure}

Loss within the inductive band is analyzed with Bode $Q$ (\(Q_\textit{Bode}\)) (Fig. \ref{BodeQ}), evaluated using the measured 1-port scattering-parameter data as \cite{jin_improved_2021}:

\begin{equation}
Q_{\mathrm{Bode}}
= \omega \left|\frac{dS_{11}}{d\omega}\right|\frac{1}{1-|S_{11}|^{2}}
\label{eq:Q_improved}
\end{equation}

\noindent
where $S_{11}(\omega)$ is the complex reflection coefficient of the resonator referenced to the measurement system impedance. $\omega = 2\pi f$ is the angular frequency, and $\left|dS_{11}/d\omega\right|$ is the magnitude of the frequency derivative of $S_{11}$. The term $1-|S_{11}|^{2}$ corresponds to the fraction of incident power delivered into the device. The final product is dimensionless, yielding a quality-factor estimate that remains well behaved in the presence of spurious modes. Utilization of frequency-dependent $Q$ is especially useful since piezoelectric power converters typically operate in a spurious-free band following the series resonance, although group-delay fluctuations from the measurement may still influence the smoothed data.

 The results are smoothed with a sliding window size of 80 data points for mitigating the impact of noise on the differential operator and shown in Fig. \ref{BodeQ} (a) - (c). Compared with the proposed structure, conventional designs exhibit significant spikes in \(Q_\textit{Bode}\) within the inductive bandwidth, attributable to high-\(Q\) in-band spurious modes. On the other hand, the proposed structure yields a smoother in-band \(Q_\textit{Bode}\), providing a wide operating region for the power converter with low loss. 
 
The extracted key performance parameters, including $k^2$, 3-dB \(Q\) at series resonance ($Q_s$), maximum \(Q_\textit{Bode}\) in-band ($Q_{max}$), and the FoM calculated using $Q_{max}$ (\(FoM_\textit{max}\)) are summarized in Table \ref{tab2}, indicating a high \(FoM_\textit{max}\) of 1548. Note that the rectangular reference achieves the highest $Q_{\max}$ of 7924; however, this peak occurs at a spurious-mode resonance within the inductive band, making it inaccessible for converter operation. In contrast, the proposed design's $Q_{\max}$ of 5230 occurs within a spurious-free region and is therefore practically usable. \(Q_s\) follows the same pattern, as \(f_s\) of the reference devices coincides with spurious modes.

\begin{table} [!t]
\caption{Extracted Key Performance Parameters}
\label{tab2}
\centering
\renewcommand{\arraystretch}{1.3}
\begin{threeparttable}
\begin{tabular}{M{60pt}
                M{25pt}
                M{20pt}
                M{20pt}
                 M{20pt}
                M{25pt}}
\hline\hline
\text{Design} & \text{$f_s$ (MHz)} & \text{$k^2$ (\%)} & \text{$Q_s$} & \text{$Q_{max}$} & $\mathrm{FoM}_{\max}$ \\
\hline
Rectangular Ref. & 10.23 & 31.8 & 5100 & 7924 & 2520 \\
Circular Ref. & 10.26 & 31.6 & 4560 & 4975 & 1572 \\
\textbf{Proposed Design} & \textbf{10.14} & \textbf{29.6} & \textbf{3750} & \textbf{5230} & \textbf{1548} \\
\hline\hline
\end{tabular}%
\begin{tablenotes}
\footnotesize
\item[] \centering *Reference-device \(Q_s\) are affected by spurious modes.
\end{tablenotes}
\end{threeparttable}
\end{table}

\begin{figure}[!t]
\centerline{\includegraphics[width=\columnwidth]{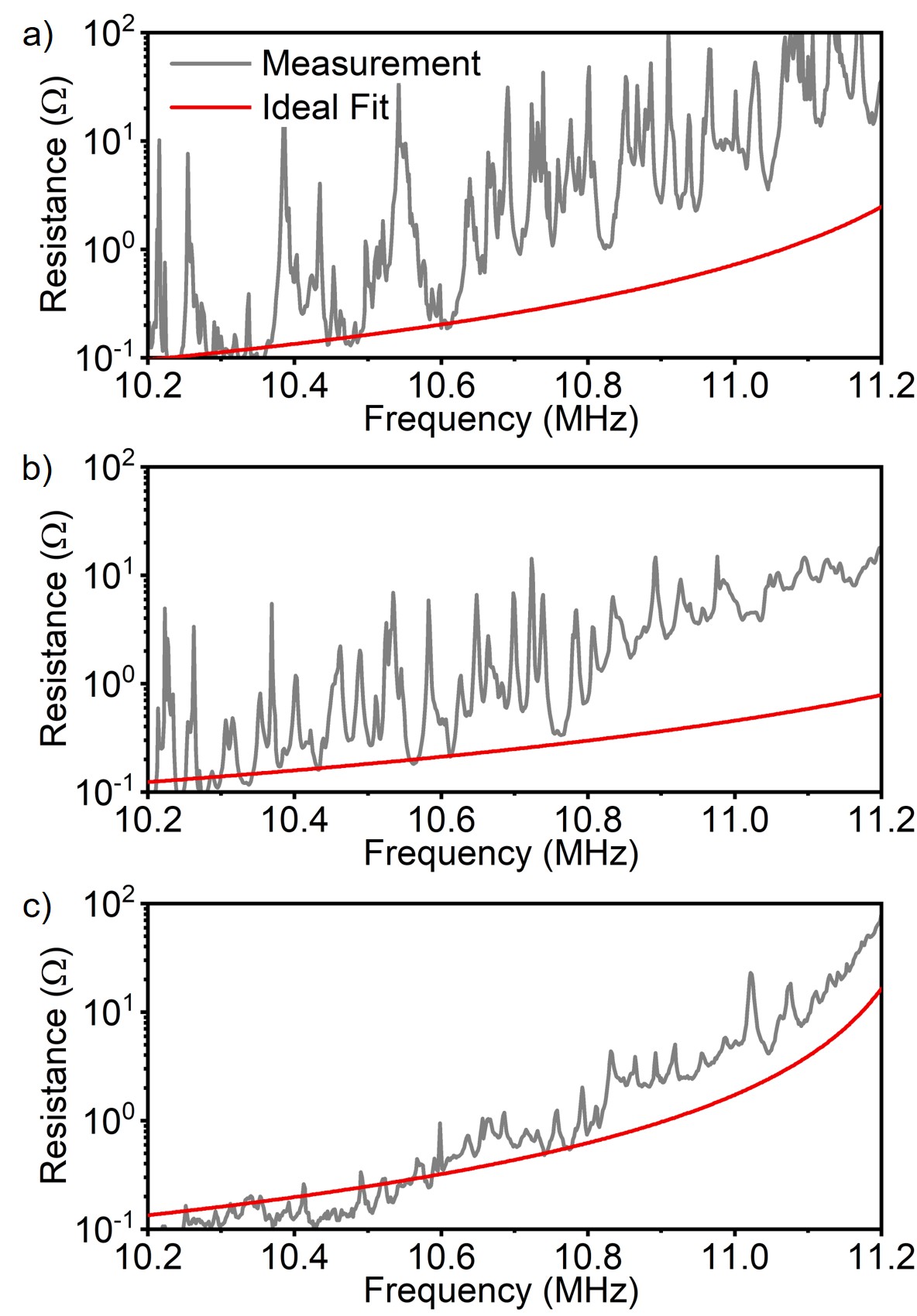}}
\caption{Zoomed in measured resistance curves and an ideal spurious-free case obtained with the BVD model for (a) a conventional rectangular device, (b) a conventional circular device, and (c) the proposed grounded-ring structure. The grounded-ring resonator shows the least deviation from the ideal case.}
\label{ZoomResistance}
\end{figure}

The efficacy of the spurious-mode suppression mechanism is further validated by comparing the measurements with an idealized single-branch BVD equivalent circuit, which yields an idealized single-mode inductive response. As shown in Fig. \ref{ZoomResistance} (a)-(c), the reference designs deviate significantly from the ideal BVD response due to spurious modes. The difference is apparent: the grounded-ring device most closely follows the ideal single-branch BVD behavior and shows the clearest suppression of spurious features.

\subsection{Comparison to Prior Works}

The key performance parameters are summarized in Table~\ref{tab2}, where the grounded-ring resonator achieves a high $\mathrm{FoM}_{\max}$ of 1548 with its high \(k^2\) and a high measured  $Q_\textit{max}$. Compared with the SoA reported in Table~\ref{tab1}, the achieved result is particularly beneficial for piezoelectric power conversion: prior demonstrations have shown high FoM only at lower frequencies or exhibited significant spurious modes. In contrast, the proposed structure exhibits a high resonance frequency and a spurious-free response over a wide frequency range, thereby enabling the achieved FoM to be practically leveraged for high-power-density piezoelectric power conversion.

Large-signal characterization (power handling, nonlinearity, thermal behavior) and converter-level integration have been reported separately in \cite{stolt_spurious-free_2024, daniel_nonlinear_2024}; this work focuses on the acoustic metrics that most directly determine usable converter bandwidth and efficiency: $k^2$, $Q$, and spurious-mode suppression.

\subsection{Laser Doppler Vibrometry Measurement}

\begin{figure*}[!t]
\centerline{\includegraphics[width=\textwidth]{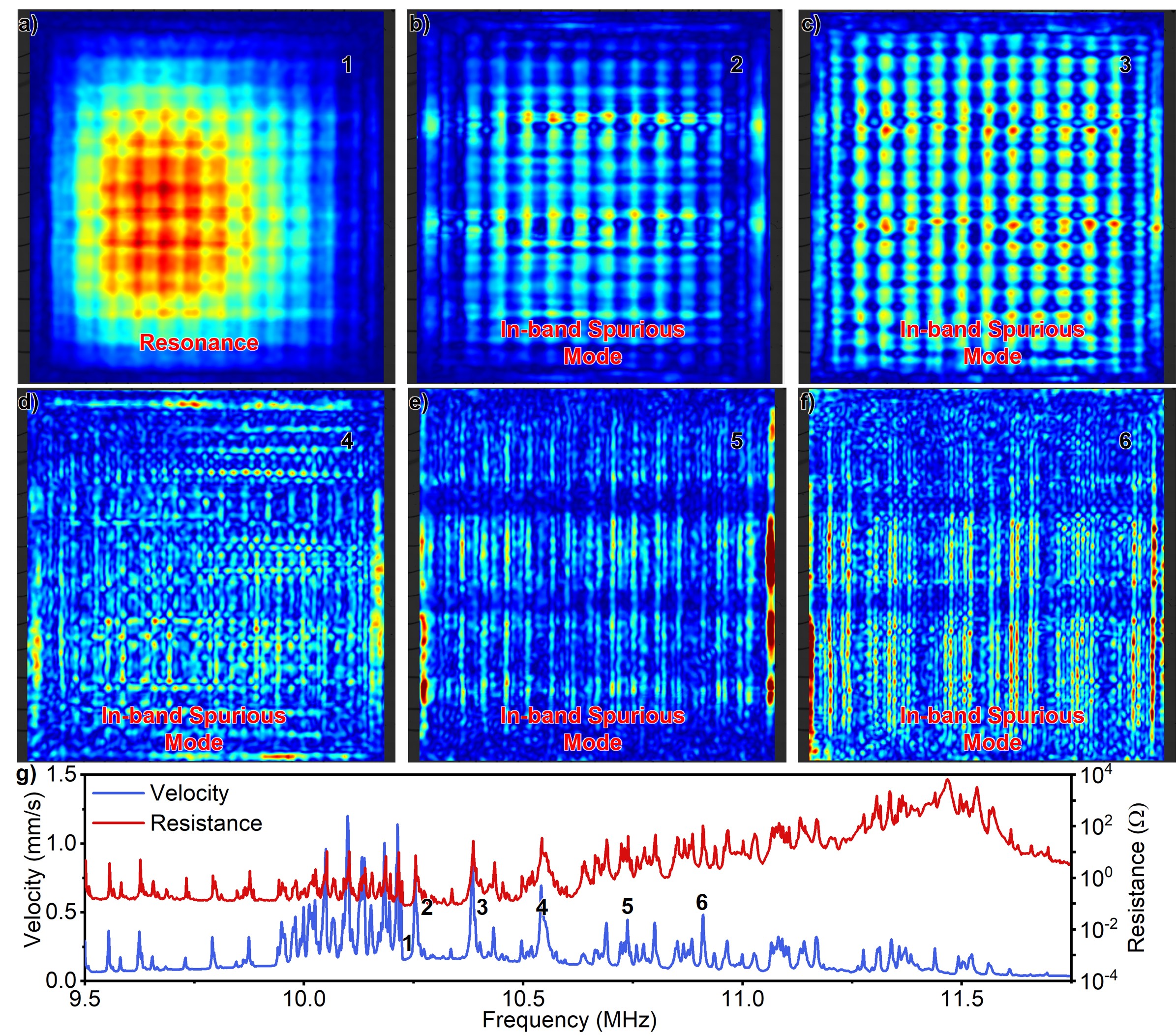}}
\caption{LDV characterization of the rectangular reference TE-mode LN BAW resonator (without a grounded ring). (a)-(f) Measured \(v_z\) deflection profiles (colorbar: \(v_z\) magnitude in mm/s, blue = minimum, red = maximum) at the six frequencies labeled 1-6. (g) \(v_z\) amplitude (blue, left axis) overlaid with the measured electrical resistance (red, right axis), showing a high density of in-band spurious modes across the inductive operating region.}
\label{rectrefldv}
\end{figure*}

\begin{figure*}[!t]
\centerline{\includegraphics[width=\textwidth]{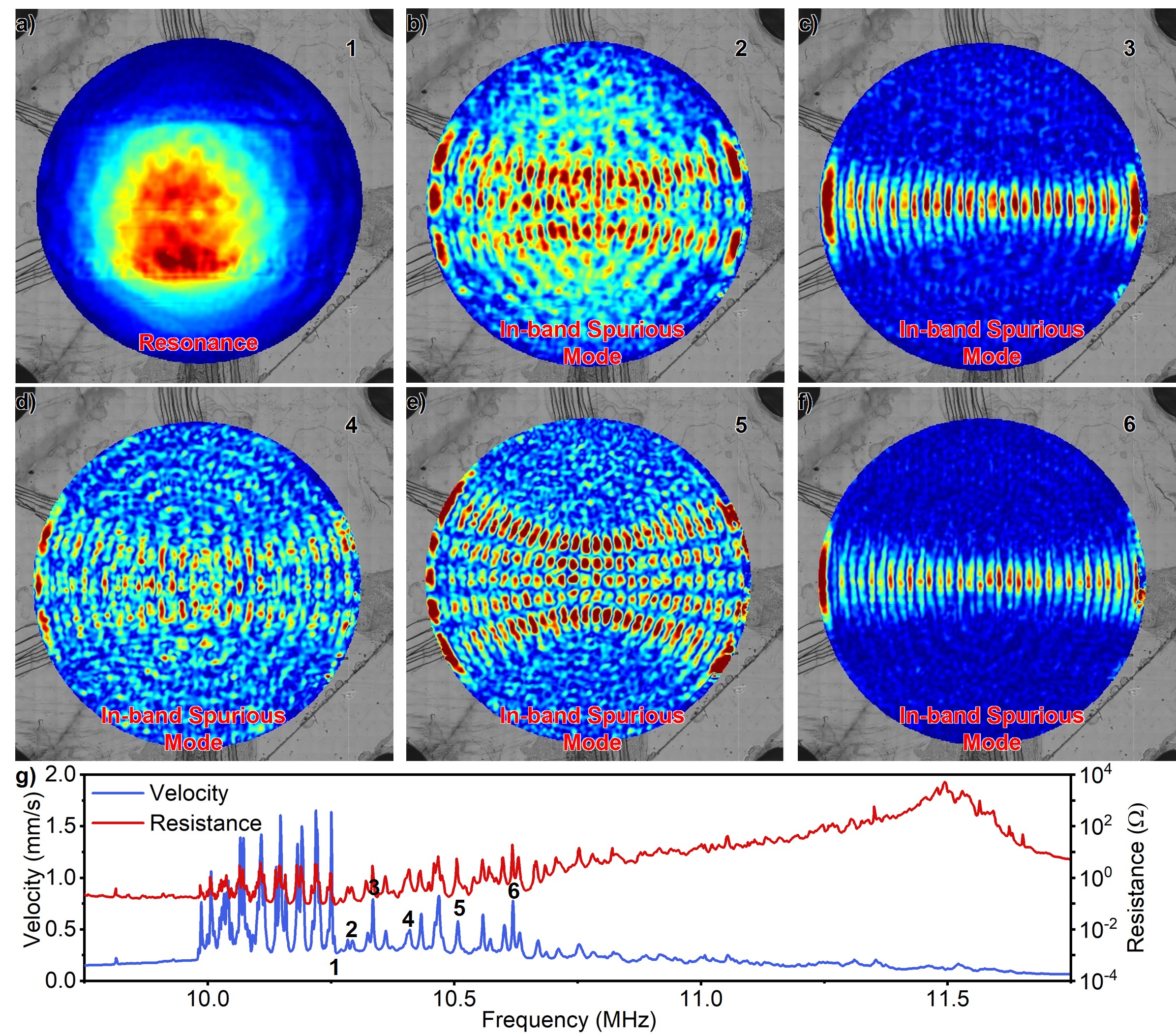}}
\caption{LDV characterization of the circular reference TE-mode LN BAW resonator (without a grounded ring). (a)-(f) Measured \(v_z\) deflection profiles (colorbar: \(v_z\) magnitude in mm/s, blue = minimum, red = maximum) at the six frequencies labeled 1-6. (g) \(v_z\) amplitude (blue, left axis) overlaid with the measured electrical resistance (red, right axis), showing in-band spurious modes.}
\label{circrefldv}
\end{figure*}

\begin{figure*}[!t]
\centerline{\includegraphics[width=\textwidth]{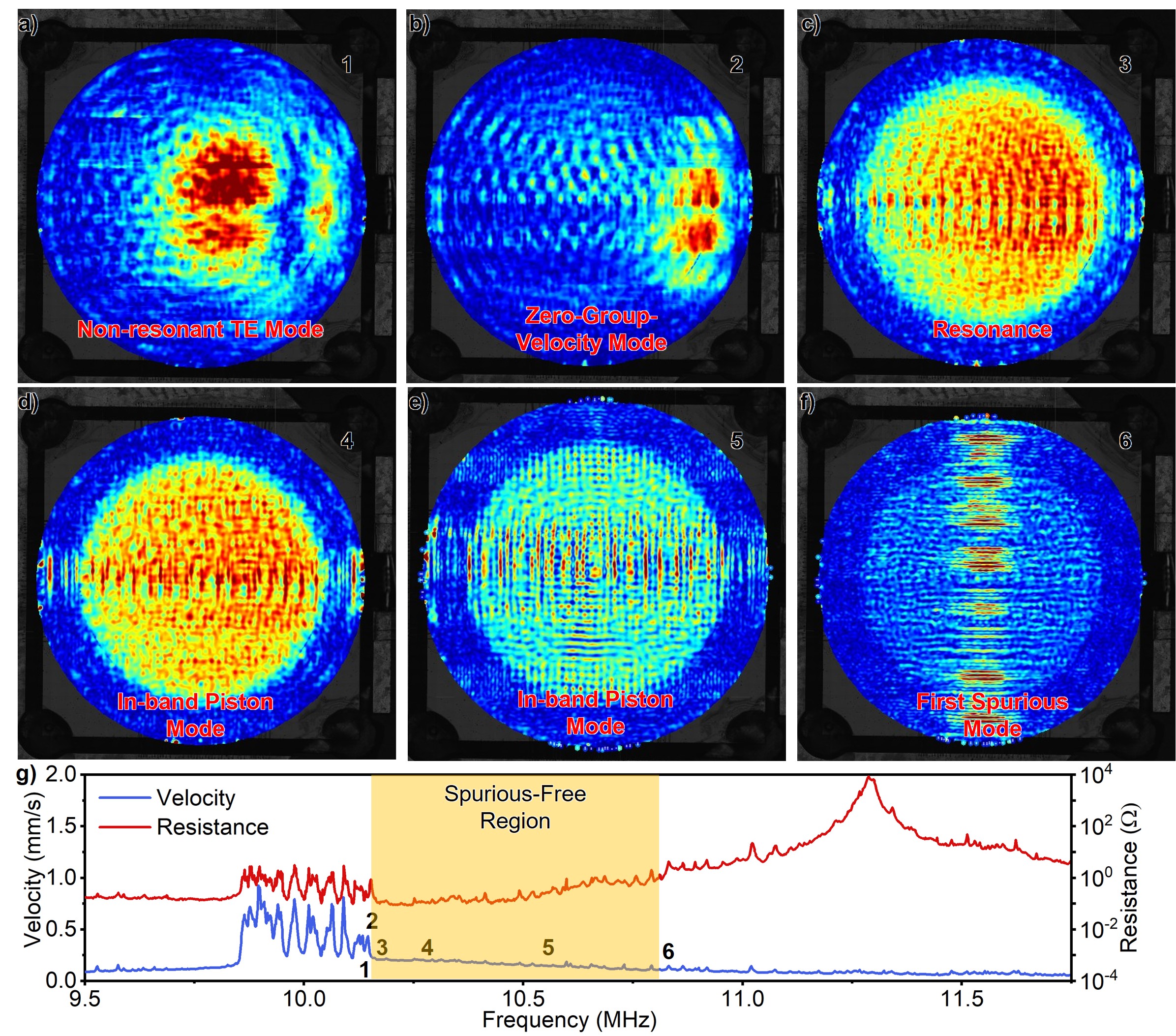}}
\caption{LDV characterization of the proposed grounded-ring TE-mode LN BAW resonator. (a)-(f) Measured \(v_z\) deflection profiles (colorbar: \(v_z\) magnitude in mm/s, blue = minimum, red = maximum) at the six frequencies labeled 1-6. (g) \(v_z\) amplitude (blue, left axis) overlaid with the measured electrical resistance (red, right axis), demonstrating spurious-mode suppression across the inductive band.}
\label{circringldv}
\end{figure*}

To provide insight into the operating mechanism of the proposed grounded-ring structure, the rectangular reference, circular reference, and circular grounded-ring devices were measured using a Polytec laser Doppler vibrometer (LDV), as shown in Figs. \ref{rectrefldv}, \ref{circrefldv}, and \ref{circringldv}, respectively. A chirp signal was applied to the devices, and the resulting performance was extracted for the piezoelectric resonators. LDV measurements were taken using a Polytec MSA-100-3D system in 1D mode. The system's XY positioning stage scanned over 130,000 measurement locations across the device surface. Scan points were defined in a square mesh pattern with 45 \(\upmu\)m lateral spacing in the in-plane directions. During the scan, the system's built-in function generator continuously excited the device with a periodic chirp signal between 9 MHz and 12.5 MHz. The out-of-plane surface velocity responses were measured sequentially via the MSA's emitted laser at each pre-defined location. Because the generator signal was also used as a reference, the relative phases and magnitudes between the points could be captured. The result is a mapping of the device's complex out-of-plane velocity ($v_z$) response across its entire surface, enabling visualization and animation of deflection shapes for any given frequency bin within the measured band.

The LDV measurements show out-of-plane deflection profiles that map $v_z$ of the reference and grounded-ring resonators throughout the inductive band, which are representative of the mechanical energy distribution within the resonators. Fig. \ref{rectrefldv}, Fig. \ref{circrefldv} and Fig. \ref{circringldv} (g) show the extracted spatially averaged $v_z$ spectrum and the measured resistance of the resonator. $v_z$ is averaged over the device surface to compare with electrical data measured using the VNA, which reflects the total motional current induced by the piezoelectric effect. Spatially averaged $v_z$ amplitude correlates closely with the VNA-measured resistance, with velocity peaks aligning with the spurious modes, while its complex average can be correlated with the electrical admittance.

The rectangular reference design [Fig.~\ref{rectrefldv}] shows a TE-mode resonance at 10.24~MHz followed by dense high-order standing waves across the entire inductive band, confirming that spurious modes severely restrict operational bandwidth. The circular reference design shows a similar TE-mode resonance at 10.25 MHz [Fig. \ref{circrefldv} (a)], and is followed by a number of lateral overtones which primarily limit the resonator's low-impedance inductive region [Fig. \ref{circrefldv} (b)-(f))]. However, compared with the rectangular reference, the circular structure exhibits a significantly lower density of spurious modes overall. This difference is attributed to the significant in-plane anisotropy in LN, which inhibits the formation of lateral standing waves in an isotropic geometry \cite{kuznetsova_power_2008, touhami_piezoelectric_2021}. Hence, the remaining spurious modes in the circular reference design are primarily lateral TE-mode overtones, which do not exhibit significant in-plane anisotropy, and the majority of spurious modes with dominant lateral components are suppressed by the circular geometry. Nonetheless, the LDV measurement shows that the inductive band in the circular reference design is constrained by significant spurious modes between 10.25~MHz and 10.7~MHz, which comprise the resonator's low-impedance inductive region, severely limiting its use for piezoelectric power conversion despite its high FoM.

On the other hand, the proposed structure demonstrates a major shift in the excited modal profiles, as seen in Fig. \ref{circringldv} (a)-(f). The response shifts from a conventional TE mode to a piston-like velocity profile at resonance, which is maintained within the inductive bandwidth. The TE mode can still be observed as a non-resonant tone at a frequency below series resonance, superimposed with an additional localized velocity antinode within the narrow gap [Fig. \ref{circringldv} (a)], followed by a dedicated resonance in the gap, which is a major spurious mode immediately prior to \(f_s\) [Fig. \ref{circringldv} (b)]. This gap resonance falls below $f_s$ and therefore lies outside the converter's inductive operating band; it does not degrade the converter's performance. In contrast to the reference design, the proposed structure excites a piston-like modal profile rather than a conventional TE mode at resonance [Fig. \ref{circringldv} (c)]. Importantly, Fig. \ref{circringldv} (d) and (e) show that lateral TE-mode overtones are suppressed, and a piston-like mode is the representative mode-shape within the inductive bandwidth. Conventional spurious modes, such as those seen in Fig. \ref{circrefldv} (f), begin to obtain small coupling only when approaching \(f_p\), which is not utilized in the given piezoelectric power converter topology. It is important to note that the left-hand side of the obtained spatial profile is significantly damped due to difficulties in achieving a good reflection from parts of the device covered by the wirebonds, which can be observed in Fig. \ref{mounteddevice}.

The extracted surface deflection profiles (Fig. \ref{GroundedRing_LDVProfiles}) reveal a major difference between the two designs: a conventional design exhibits a half-wavelength in the active area at resonance, with a velocity node at the edge of the electrodes, confirming TE-mode excitation. In contrast, the grounded-ring device shows uniform displacement over the entire active area and decays only in the gap and grounded-ring regions, characteristic of a piston mode, which has been shown to achieve spurious-mode suppression \cite{kaitila_spurious_2003, thalhammer_4e-3_2006, yang_piston-like_2025}.

\section{Spurious-Mode Suppression Mechanism}

\subsection{Piston Mode Operation}

\begin{figure}[!t]
\centerline{\includegraphics[width=\columnwidth]{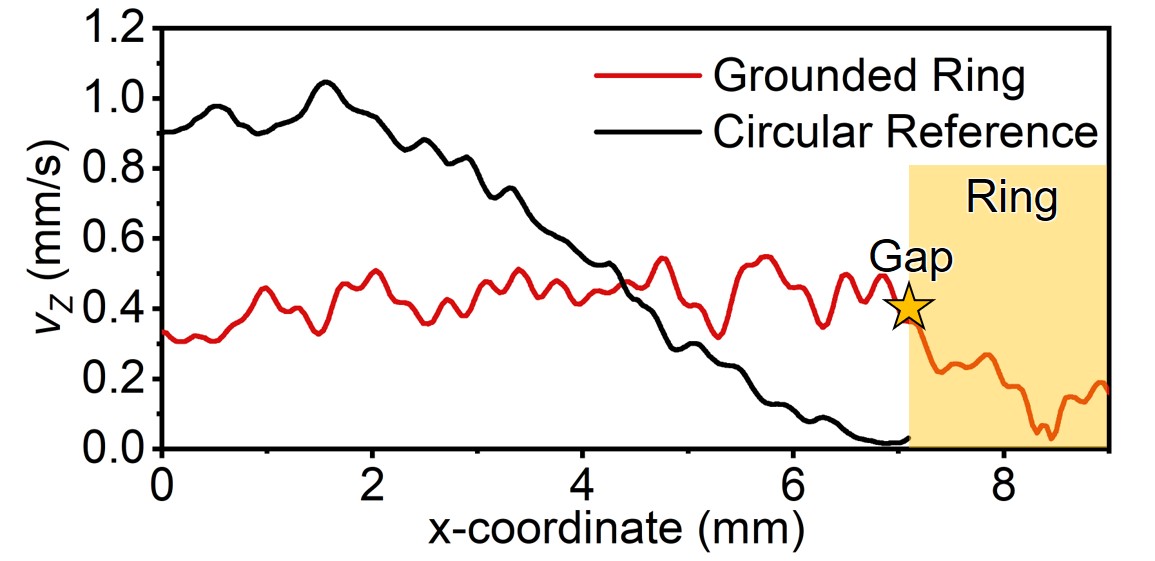}}
\caption{Comparison of measured resonant mode-shapes between the circular reference and grounded-ring designs, demonstrating a TE-mode resonance in the conventional structure, while the grounded-ring device shows a uniform piston-like \(v_z\) profile.}
\label{GroundedRing_LDVProfiles}
\end{figure}

\begin{figure}[!t]
\centerline{\includegraphics[width=\columnwidth]{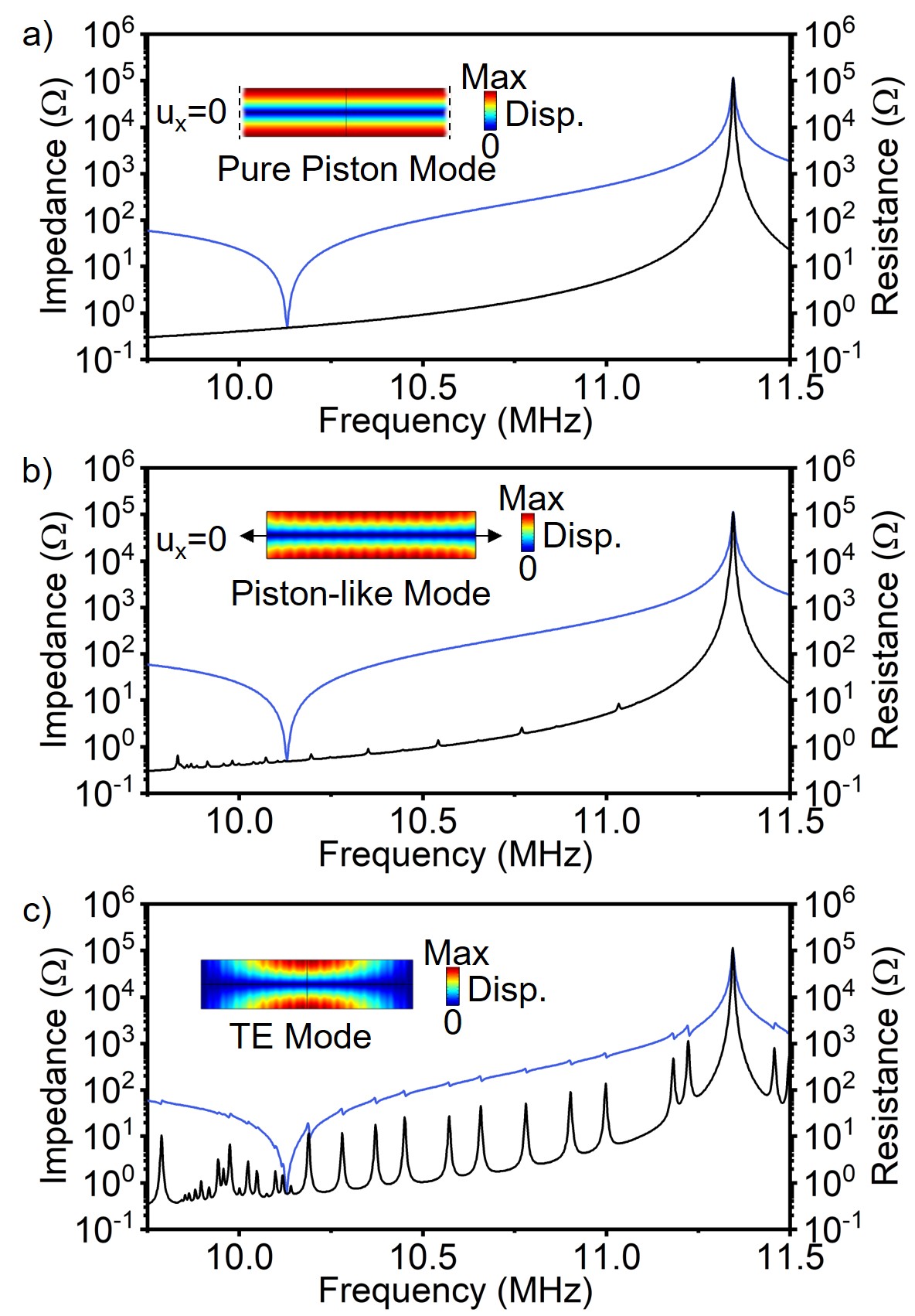}}
\caption{Simulated impedance and resistance profiles from 2D cross-sectional FEA for (a) zero \(u_x\) at the lateral boundaries, leading to a pure piston mode. (b) Zero \(u_x\) at the midplane, exciting a piston-like mode. (c) A free boundary condition at the device edge, leading to the TE mode.}
\label{GroundedRing_PistonProfiles}
\end{figure}

The LDV measurements in Figs. \ref{circrefldv} and \ref{circringldv} provide direct experimental validation of the effect of the narrow gap and clarify the underlying operating mechanism of the proposed structure. The measured surface deflection from the conventional device exhibits a TE resonance, which is a special case of the first-order symmetric Lamb (S1) mode achieved when the device lateral dimensions far exceed the plate thickness. In contrast, the grounded-ring structure exhibits a different special case of S1, the piston mode. The ideal piston mode is characterized by uniform out-of-plane displacement at the piezoelectric device surface, with a zero lateral wavenumber (\(k_x\)). This mode can be excited by imposing boundary conditions on lateral particle displacement (\(u_x\)), and can be shown to be intrinsically spurious-free [Fig. \ref{GroundedRing_PistonProfiles} (a)] \cite{kaitila_spurious_2003, thalhammer_4e-3_2006, yang_piston-like_2025}. While a pure piston mode requires \(u_x\) $= 0$ across the entire lateral boundary, a partial approximation is achieved when \(u_x\) is constrained at discrete locations along the boundary. For instance, Fig. \ref{GroundedRing_PistonProfiles} (b) shows a piston-like mode excited by enforcing \(u_x\) $= 0$ near the mid-plane (half-thickness) of the piezoelectric plate. This boundary condition suppresses a significant portion of the relevant spurious modes, especially when compared with a device operating under free boundary conditions [Fig. \ref{GroundedRing_PistonProfiles} (c)], which represents the benchmarked reference design without a grounded ring. A comparison of these simulations with the measured displacement profiles in Fig. \ref{GroundedRing_LDVProfiles} indicates that the grounded-ring structure excites a piston-like mode, and the spurious-mode suppression mechanism is achieved by the grounded ring altering the effective boundary conditions of the device.

\subsection{Lamb Wave Dispersion in LN}

\begin{figure}[!t]
\centerline{\includegraphics[width=\columnwidth] {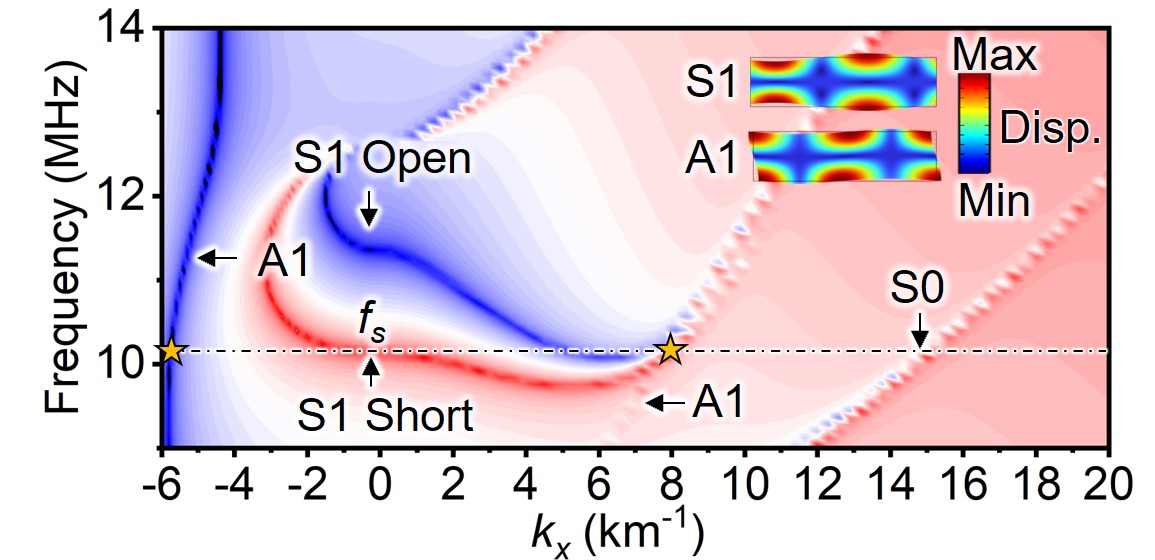}}
\caption{Dispersion analysis of the proposed structure. The obtained solutions are color-coded as blue for solutions in electrically open LN and red for solutions in electrically short LN. Line brightness corresponds to the admittance magnitude obtained from the periodic boundary condition sweep: brighter regions indicate stronger resonant responses.}
\label{GroundedRing_dispersion}
\end{figure}

The proposed design relies on the dispersion characteristics of LN to alter the boundary condition at the edge of the active area and excite a piston-like mode at resonance. The associated dispersion curves of metalized LN (electrically short), and non-metalized LN (electrically open) were extracted using COMSOL FEA (Fig. \ref{GroundedRing_dispersion}). They were obtained by defining an elementary LN unit cell using a one-dimensional mesh and solving for solutions that satisfy varying periodic Floquet boundary conditions. The results in Fig. \ref{GroundedRing_dispersion} show that LN exhibits type-2 dispersion, in which the longitudinal wave velocity is less than twice the shear wave velocity, resulting in the unique dispersion characteristics of S1, e.g., negative group velocity and two distinct cutoff frequencies \cite{fattinger_optimization_2005, maksimov_bound_2006}. Hence, low-order S1 overtones are excited below \(f_s\), while the high-order S1 modes create regions of loss for the piezoelectric power converter and are the primary target for suppression.

To understand acoustic resonator behavior in the relevant low-impedance region for piezoelectric power conversion, the series resonance ($f_s$) of the LN resonator can be mapped onto the dispersion curve by drawing a horizontal line near \(k_x\) = 0, corresponding to the large lateral dimension of the active region compared with the plate thickness. This line simultaneously intersects several high-\(k_x\) S1 and first-order antisymmetric Lamb (A1) mode solutions in both electrically short and electrically open regions. Immediately below resonance, S1 in electrically open LN exhibits a real solution corresponding to a zero-group-velocity (ZGV) mode with a non-zero \(k_x\), explaining the localized resonance observed in the gap in Fig. \ref{circringldv} (b). At $f_s$, the dispersion diagram shows an intersection between high-order S1 and A1 overtones in electrically short LN at the same \(k_x\). As a result, Fig. \ref{GroundedRing_dispersion} indicates that the wave behavior in the grounded-ring structure at $f_s$ is complex, with multiple real solutions simultaneously present in all three regions of the acoustic resonator.

\begin{figure}[!t]
\centerline{\includegraphics[width=0.95\columnwidth]{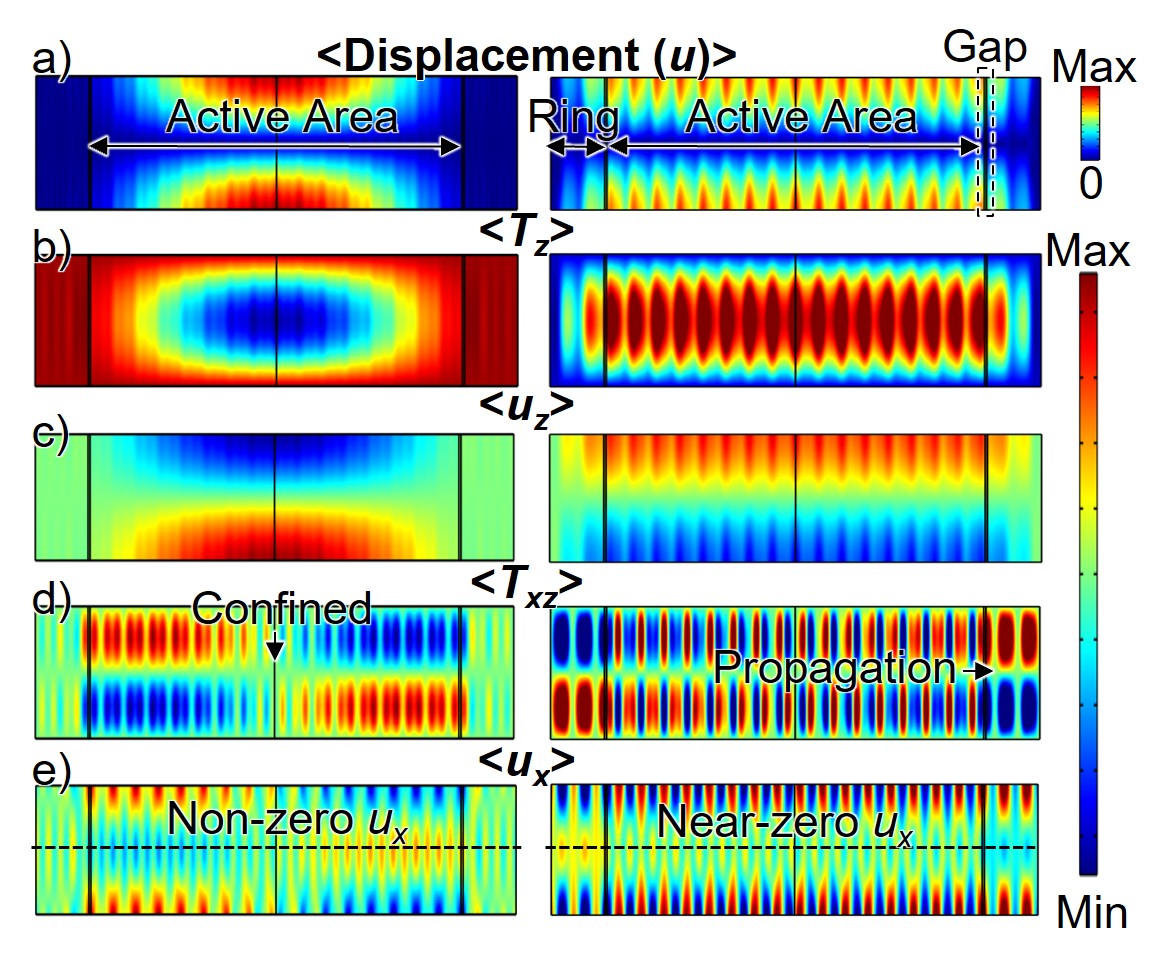}}
\caption{Simulated (a) displacement, (b) out-of-plane stress, (c) out-of-plane displacement, (d) shear stress, and (e) in-plane displacement of the reference (left) and grounded-ring (right) resonators, highlighting the piston-like mode and distinct shear-stress and lateral-displacement profiles in the grounded-ring design. The mode-shapes of the grounded-ring structure are normalized to that of the reference device.}
\label{GroundedRing_Resonance_ModeShapeInspection}
\end{figure}

Based on these observations, \(k_x\) for all relevant intersections in Fig. \ref{GroundedRing_dispersion} are extracted for analysis. These intersections yield three distinct characteristic lateral wavelengths (\(\lambda_x\)) of \(\lambda_x\) $= 1.26$ mm for S1 in electrically open LN with \(k_x\)$=5000\,\mathrm{m^{-1}}$, \(\lambda_x = 790\,\upmu\mathrm{m}\) for both S1 and A1 overtones in electrically short LN with \(k_x\)$=7945\,\mathrm{m^{-1}}$, and an evanescent decay length of 172 \(\upmu\)m for A1 in electrically open LN with \(k_x\)$=-5800\,\mathrm{m^{-1}}$.  These values are specific to a 0.3 mm thick LN plate (dimensions in Fig. \ref{GroundedRing_schematic}), and scale accordingly for other operating frequencies. The following analysis will show that these parameters play a key role in determining the optimal geometry of the grounded-ring resonator.

\subsection{Working Mechanism Analysis}

\begin{figure}[!t]
\centerline{\includegraphics[width=\columnwidth]{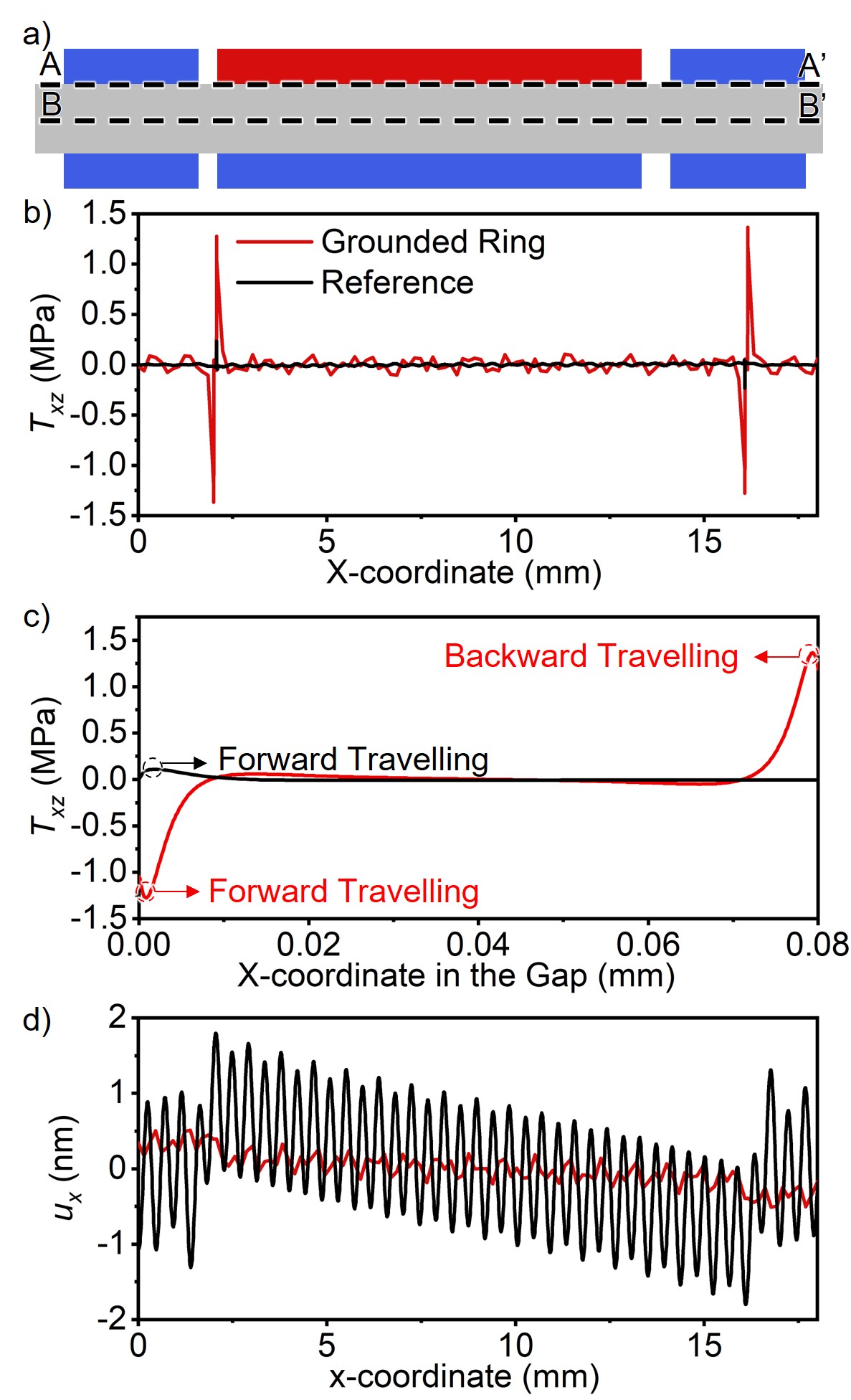}}
\caption{(a) Cut-line locations for data extraction; a reference device without the grounded ring is used for comparison. (b) \(T_\textit{xz}\) along A-A', showing opposing-polarity antinodes at the gap interfaces. (c) Zoomed \(T_\textit{xz}\) profile within the gap, highlighting antinodes excited at both the active area/gap and the gap/ring interface with opposing polarities traveling in opposite directions. (d) Lateral displacement along B-B', highlighting suppressed \(u_x\) in the grounded-ring structure.}
\label{GroundedRing_Txy_ux_inspection}
\end{figure}

Frequency domain FEA was used to compare mode shapes of the conventional and the grounded-ring designs with dimensions from Fig. \ref{GroundedRing_schematic}, shown in Fig. \ref{GroundedRing_Resonance_ModeShapeInspection} (a) - (e). The reference structure excites a conventional TE mode with standard out-of-plane stress (\(T_z\)) and out-of-plane displacement (\(u_z\)) profiles. In contrast, the grounded-ring structure shows a piston-like mode resembling that in Fig. \ref{GroundedRing_PistonProfiles} (b) [Fig. \ref{GroundedRing_Resonance_ModeShapeInspection} (a)-(c)].  The critical difference emerges in the shear stress (\(T_\textit{xz}\)) and lateral displacement (\(u_x\)) profiles [Fig. \ref{GroundedRing_Resonance_ModeShapeInspection} (d) and (e)]. In a conventional geometry, these profiles exhibit evanescent decay of acoustic energy outside of the active area. However, the grounded-ring structure exhibits acoustic coupling between the active area and the grounded ring, with propagating mode shapes in both regions. To illustrate how the coupling occurs, cut-line data was extracted from both geometries, shown in Fig. \ref{GroundedRing_Txy_ux_inspection} (a)-(d). Inspection of the \(T_\textit{xz}\) profile at the device surface reveals that the grounded-ring structure exhibits exponentially decaying stress antinodes with opposing polarities at the active area/gap and the gap/grounded-ring interface, while the conventional structure shows minimal \(T_\textit{xz}\) [Fig. \ref{GroundedRing_Txy_ux_inspection} (b) and (c)]. Fig. \ref{GroundedRing_Txy_ux_inspection} (d) shows that \(u_x\) is suppressed throughout the entire structure, indicating that the propagating \(T_\textit{xz}\) profile in the active and grounded-ring regions seen in Fig.\ref{GroundedRing_Resonance_ModeShapeInspection} (d) may be integral to the working mechanism. This indicates that the grounded-ring geometry introduces boundary conditions akin to those seen in Fig. \ref{GroundedRing_PistonProfiles} (b), which enable intrinsic spurious-mode suppression by exciting a piston-like mode.

\begin{figure}[!t]
\centerline{\includegraphics[width=\columnwidth]{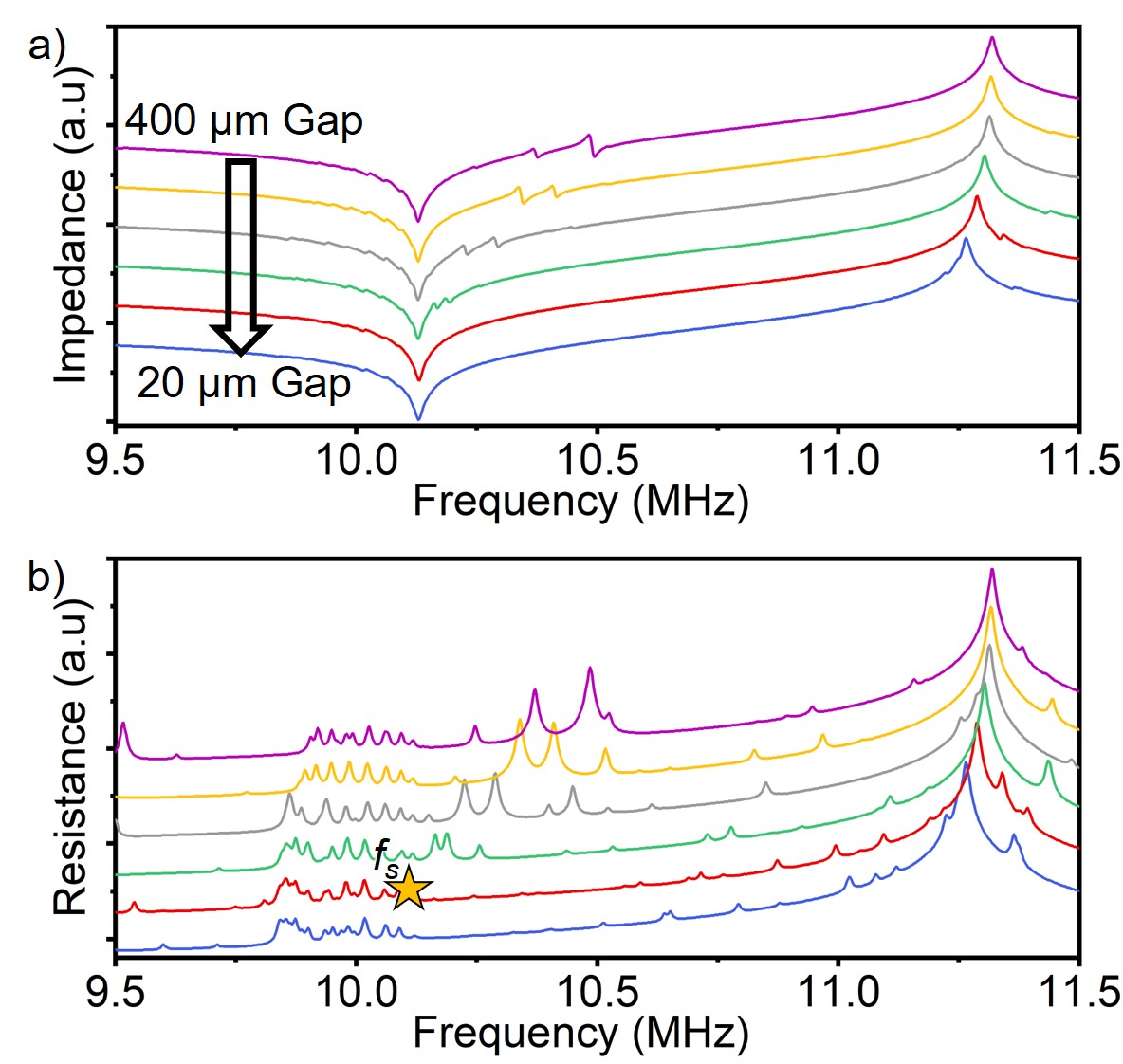}}
\caption{Simulated impedance and resistance curves for the grounded-ring resonator with parameters seen in Fig. \ref{GroundedRing_schematic} while sweeping the gap width from 20 \(\upmu\)m to 400 \(\upmu\)m, highlighting the spurious-mode suppression with gap widths below the decay length (bottom three curves), and the excitation of lower-order S1 overtones when the gap width exceeds the decay length. Impedance and resistance curves are vertically offset for clarity (arbitrary units on the ordinate). The parameter utilized in the proposed design is denoted by a star.}
\label{GroundedRing_GapSweep}
\end{figure}

Prior studies on Lamb wave dispersion confirm that the observed intersection of S1 and A1 in Fig. \ref{GroundedRing_dispersion} can impose constraints on particle displacement due to their interaction throughout the geometry \cite{mindlin_mathematical_1957, veres_crossing_2014, pilarski_remarks_1993, chati_maxima_2011, freedman_variation_1990, lawrie_edge_2012}. Meanwhile, Fig. \ref{GroundedRing_PistonProfiles} shows that constraints on \(u_x\) can excite a piston-like mode. Accordingly, the ability of the grounded-ring geometry to support co-propagation of such waves may be the key to the spurious-mode suppression mechanism, with the narrow gap providing the evanescent coupling path, as indicated by the exponentially decaying \(T_\textit{xz}\) profile in Fig. \ref{GroundedRing_Txy_ux_inspection} (b)-(c). As a result, the proposed framework dictates that geometric parameter selection is key to enabling spurious-mode suppression, and can be guided solely by the Lamb wave dispersion in Fig. \ref{GroundedRing_dispersion}.

\subsection{Geometric Parameter Selection}

\begin{figure}[!t]
\centerline{\includegraphics[width=\columnwidth]{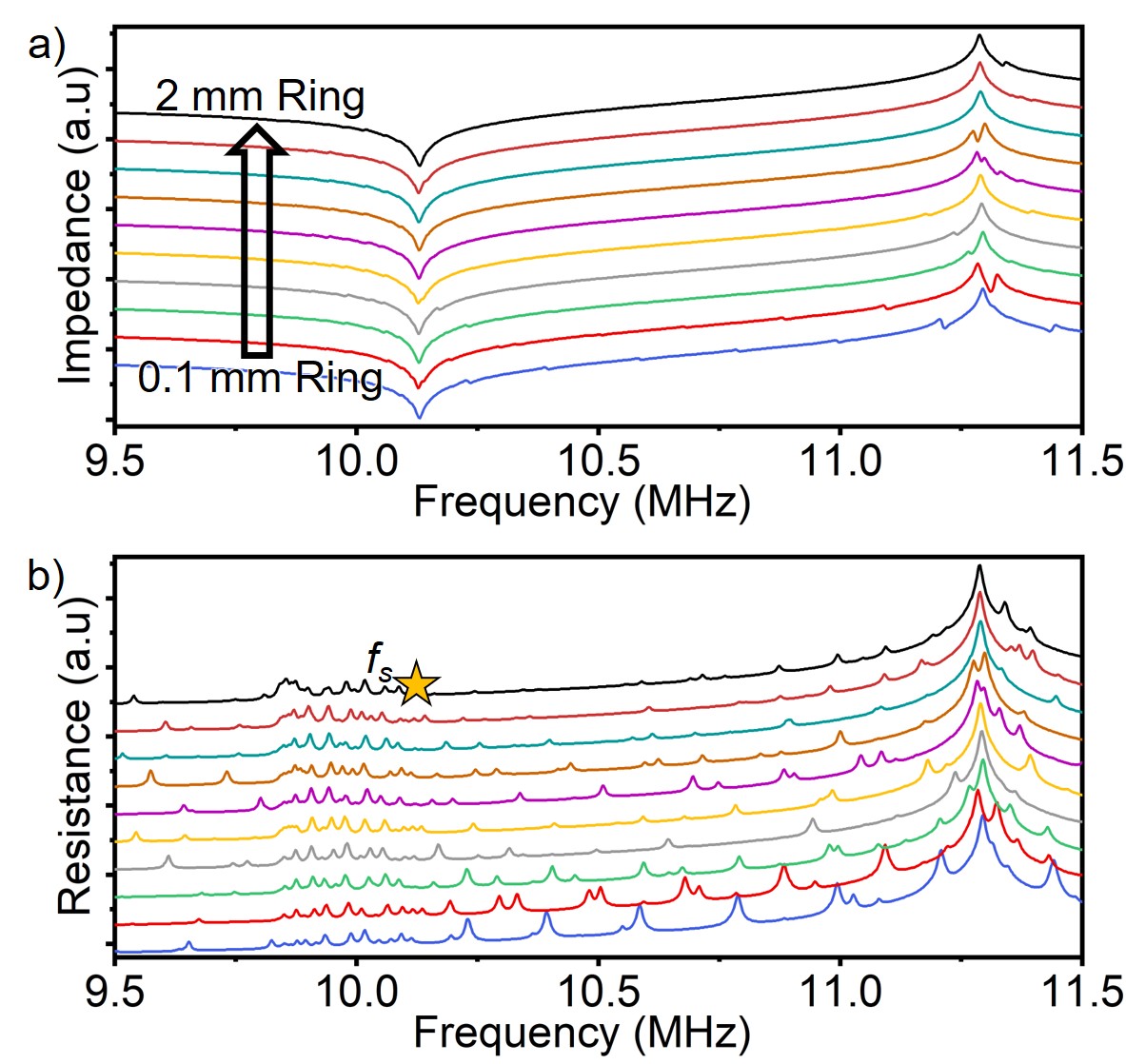}}
\caption{Simulated impedance and resistance curves for the grounded-ring resonator with parameters seen in Fig. \ref{GroundedRing_schematic} while sweeping the grounded-ring width from 0.1 mm to 2 mm, indicating that the spurious-mode suppression is effective when the ring width exceeds the spurious S1 overtone \(\lambda_x\) of 790 \(\upmu\)m (top three curves). Impedance and resistance curves are vertically offset for clarity (arbitrary units on the ordinate). The parameter utilized in the proposed design is denoted by a star.}
\label{GroundedRing_RingSweep}
\end{figure}

A closer look at the wave behavior in the narrow gap informs the optimal choice of the geometric design parameters to enhance spurious-mode suppression in the grounded-ring resonator. Analysis of the \(T_\textit{xz}\) profile at the gap surface revealed exponentially decaying solutions with an extracted decay length of 172 \(\upmu\)m. Its physical significance is reinforced by an impedance simulation with a varying gap width, indicating that spurious-mode suppression is effective only when the gap width is below the decay length [Fig. \ref{GroundedRing_GapSweep} (a) and (b)]. For instance, gap widths up to 180 \(\upmu\)m, represented by the lower three curves in Fig. \ref{GroundedRing_GapSweep}, exhibit a similar spurious-free profile near \(f_s\). Past this value, the spurious-free region shrinks significantly, primarily due to the high-coupling S1 overtones.

The grounded-ring width is another key consideration for spurious-mode suppression. A parametric sweep reveals that for ring widths below 790 \(\upmu\)m, the design is ineffective even when the gap width is below the decay length [Fig. \ref{GroundedRing_RingSweep} (a) and (b)]. This threshold corresponds with \(\lambda_x\) of the propagating \(T_\textit{xz}\) profile seen in Fig. \ref{GroundedRing_Resonance_ModeShapeInspection} (d). Accordingly, like the gap width, the ring width follows from the Lamb wave dispersion in LN. The ring must be wide enough to support the S1 and A1 overtones whose co-propagation is proposed to constrain \(u_x\).

Spurious-mode suppression exhibits frequency-dependent behavior within the inductive band. Figs. \ref{GroundedRing_GapSweep} and \ref{GroundedRing_RingSweep} show that with non-optimal geometric parameters, S1 overtones near \(f_s\) obtain coupling while higher-order overtones towards \(f_p\) remain suppressed. This is explained by Fig. \ref{GroundedRing_dispersion}, where the decay length of the evanescent A1 solution is shown to range from 172 \(\upmu\)m to 200 \(\upmu\)m, and \(\lambda_x\) of the S1 and A1 overtones is shown to range from 0.8 mm to 0.6 mm. Hence, with the same design parameters, S1 overtones near \(f_s\) are the most difficult to suppress, which is the key operational region for a piezoelectric power converter. On the other hand, even with the optimal geometric parameters, minor resistance spikes remain, especially near \(f_p\). These minor excited spurious modes from the 2D FEA are representative of the rectangular structure seen in Fig. \ref{rectrefldv}, which was shown to couple with a variety of modes that become suppressed in a circular geometry [Fig. \ref{circrefldv}]. Hence, they do not show up in the 3D FEA in Fig. \ref{GroundedRing_ImpedanceSims} (c) and the measurement in Fig. \ref{impedancemeasurements} (c).

The geometric design parameters specified here all scale with the plate thickness ($t$) (as governed by the Lamb wave dispersion relation). Hence, the optimal gap and ring widths scale proportionally. For example, halving the plate thickness to 150~$\upmu$m for $\sim$20~MHz operation would approximately halve the required gap and ring dimensions. Thus, the grounded-ring design framework can be scaled to the desired operating frequency to suppress spurious modes.

\subsection{Remaining Limitations and Future Development}

\begin{figure}[!t]
\centerline{\includegraphics[width=\columnwidth]{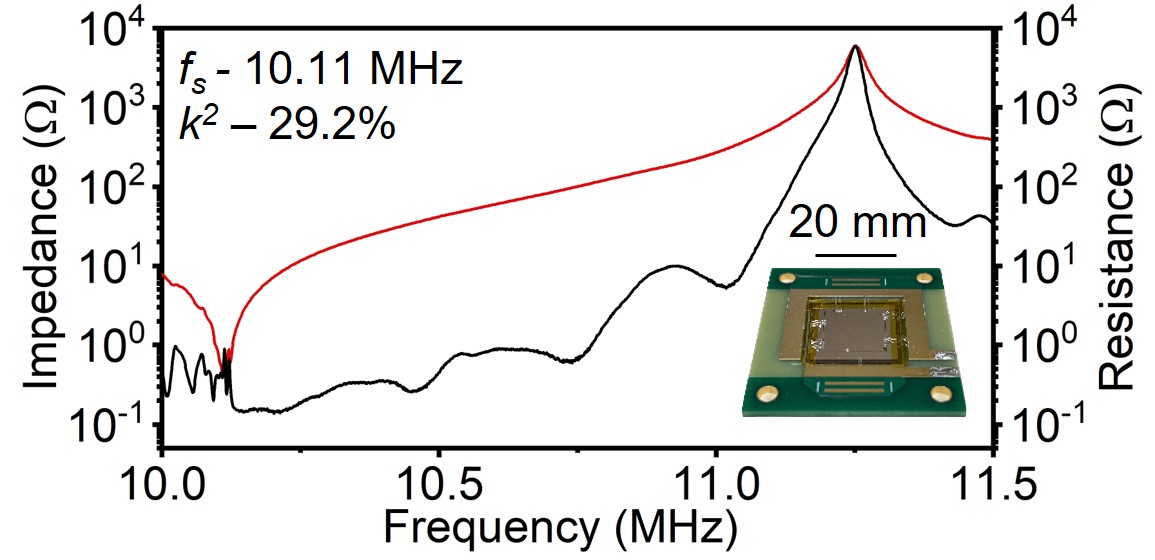}}
\caption{Impedance and resistance of a rectangular device with a grounded ring, indicating the ability of the grounded ring to introduce spurious-mode suppression regardless of the geometry. An optical image of the rectangular grounded-ring device is inset.}
\label{GroundedRing_RectangularRing}
\end{figure}

The proposed working mechanism of the grounded-ring architecture diverges significantly from the conventional spurious-mode suppression techniques reported in the literature. Prior methods rely on a single Lamb mode branch to inhibit wave propagation. In contrast, the presented design framework attributes spurious-mode suppression to the co-propagation of two Lamb waves of opposing symmetry, which establish boundary conditions fit for a piston-like mode. Although this mechanism is supported by both simulated and measured results, much of the dispersion and wave propagation analysis remains qualitative. A quantitative treatment of the interface matching conditions, such as the characterization of the wave transmission and reflection behavior across the non-metalized gap, is therefore necessary to establish the spurious-mode suppression mechanism.

Beyond the circular grounded-ring design presented in this paper, the grounded-ring approach is expected to extend to other geometries and materials. While the measurements demonstrate the inherent spurious-mode suppression ability of a circular geometry, the architecture is not restricted to circular electrodes. For example, Fig.~\ref{GroundedRing_RectangularRing} shows comparable suppression in a rectangular grounded-ring device. Hence, polygonal active areas could be considered, and the geometries between the active area and the ring could be mismatched, though a detailed treatment of these cases is beyond the scope of this work. Spurious-mode suppression via a grounded ring has also been successfully applied to LT-TE\cite{yao_lithium_2025} and AlN-TE\cite{yao_single-crystal_2026} resonators (Table~\ref{tab1}), confirming that the dispersion-guided design framework generalizes across piezoelectric material platforms.

\section{Conclusion}

In this work, we presented a grounded-ring electrode architecture for a spurious-free TE-mode LN BAW resonator targeted at piezoelectric power conversion. By introducing a grounded ring separated from the active region by a narrow gap, the proposed structure modifies the device's effective boundary conditions, suppresses lateral overtones, and preserves strong electromechanical transduction. The measured device achieved a spurious-free response at 10.14 MHz with \(k^2=29.6\%\), a maximum in-band \(Q_\textit{Bode}\) of 5230, and an FoM of 1548.

The operating mechanism was supported by both simulation and LDV characterization, showing that the grounded-ring structure promotes a piston-like modal response across the usable inductive band. This behavior enables the resonator's high \(Q\) and high \(k^2\) to be used in practice without the severe in-band loss penalties associated with spurious modes in conventional designs. These results establish the grounded-ring TE-mode resonator as a promising platform for compact, high-efficiency, and high-power-density piezoelectric power converters. More broadly, the proposed design approach offers a practical path for extending spurious-free, high-FoM acoustic resonators to higher frequencies and more demanding power-conversion applications. Beyond power conversion, the design principles demonstrated here may extend to other BAW applications where spurious-mode suppression is critical, including RF filters and sensors.

\section*{Acknowledgment}
The authors would like to thank Dr. Michael Haberman, Dr. Mark Hamilton, Dr. Jeronimo Segovia-Fernandez, and Dr. Sombuddha Chakraborty for helpful discussions.

\bibliographystyle{IEEEtran}
\bibliography{IEEESettings,thebibliography}
\end{document}